\documentclass[reprint,showpacs,twocolumn,superscriptaddress,aps]{revtex4-2}
\usepackage{amsmath,amsthm,amssymb}

\usepackage{graphicx}
\usepackage{subfig}
\usepackage{dcolumn}
\usepackage{bm}
\usepackage{caption}
\usepackage{balance}
\usepackage{bbm}

\usepackage{booktabs}
\usepackage{pgfplots}
\usepackage{graphicx}
\usepackage{lipsum,adjustbox}
\usepackage[flushleft]{threeparttable}
\usepackage{rotating}
\usepackage{booktabs}
\usepackage{multirow}
\usepackage[symbol]{footmisc}
\usepackage{stackengine}
\usepackage{tikz}
\usepackage{hyperref}
\usepackage{xurl}
\hypersetup{breaklinks=true}
\hypersetup{
    colorlinks=true,
    linkcolor=red,
    filecolor=red,
    urlcolor=red,
    citecolor=blue, 
}

\usetikzlibrary{calc, decorations.pathmorphing, decorations.pathreplacing, decorations.shapes}

\newcommand{\bra}[1]{\left\langle #1 \right|}
\newcommand{\ket}[1]{\left|#1\right\rangle}

\newcommand{\Tr}[1]{\operatorname{Tr}#1}
\makeatletter
\newcommand*{\rom}[1]{\expandafter\@slowromancap\romannumeral #1@}
\makeatother

\begin{document}


\title{Entanglement dynamics in collision models and entanglement quilts}

\author{\small Le Hu}
\email{le.hu@northwestern.edu}
\affiliation{Department of Physics and Astronomy, University of Rochester, Rochester, NY 14627, USA}
\affiliation{Institute for Quantum Studies, Chapman University, 1 University Drive, Orange, CA 92866, USA}
\affiliation{Department of Physics and Astronomy, Northwestern University, Evanston, IL 60208, USA}
\author{\small and Andrew N. Jordan}

\affiliation{The Kennedy Chair in Physics, Chapman University, Orange, CA 92866, USA}
\affiliation{Schmid College of Science and Technology, Chapman University, Orange, CA 92866, USA}
\affiliation{Institute for Quantum Studies, Chapman University, 1 University Drive, Orange, CA 92866, USA}
\affiliation{Department of Physics and Astronomy, University of Rochester, Rochester, New York 14627, USA}
\date{\today}

\begin{abstract}

We study the entanglement dynamics of a family of quantum collision models by analytically solving the pairwise concurrence for all qubit pairs. We introduce a diagrammatic method that offers an intuitive, frame-by-frame understanding of these dynamics. This allows us to monitor how a single collision affects the entanglement of the whole many-body system in some special cases. We focus on a class of models where the square of concurrence is a conserved quantity in the qubit collisions, aiding us to formulate general rules of entanglement propagation. In particular, among the multiple examples we will be showing, we identify a type of genuine multipartite entanglement, which we refer to as \textit{entanglement quilt}, where every qubit is entangled with every other qubit. We find that in some models, an entanglement quilt is hypersensitive to local excitation fluctuations: The presence of even a single excited qubit can destroy the entanglement quilts. We offer a detailed mathematical treatment on the phenomena, which can help us understand the disappearance of long-range entanglement in condensed matter systems above zero temperature. We also speculate about a possible property of the entanglement quilt: Every subsystem of it is entangled with every other subsystem.

\end{abstract}
\maketitle

\section{\label{sec:level1}Introduction}
\subsection{\label{subsec:level1}Background}
Quantum entanglement, being a fundamental concept in quantum mechanics, is essential not only for a deeper understanding on quantum physics but also for quantum information processing, such as quantum teleportation \cite{bouwmeester1997experimental}, superdense coding \cite{barreiro2008beating}, quantum computing \cite{jozsa2003role}, quantum key distribution \cite{scarani2009security}, and quantum metrology \cite{PhysRevLett.96.010401}. To understand entanglement, it is crucial to understand its dynamics, such as its formation, propagation, and decay under complex physical processes. 

Existing studies on the entanglement dynamics can be divided into few-body systems \cite{PhysRevA.65.012101, PhysRevLett.93.140404, yu2005evolution, Yonac_2007, Li_2009, Li_2020, PhysRevA.77.032342, PhysRevA.68.052306, Gennaro_2008, PhysRevLett.101.170502, PhysRevX.6.041052, MORTEZAPOUR201826, PhysRevResearch.4.033022, PhysRevA.81.052330, PhysRevA.82.032326, axioms11110589, PhysRevA.85.062323, PhysRevLett.99.160502, PhysRevA.91.012327, PhysRevLett.101.080503, PhysRevA.79.042302} and many-body systems \cite{PhysRevA.66.032110, PhysRevA.70.062304, PhysRevA.69.034304, PhysRevA.69.022304, PhysRevA.101.042324, PhysRevB.79.104428, Li_2022, PhysRevA.64.012313, PhysRevA.84.022314, PhysRevA.89.022303, PhysRevA.97.042330, PhysRevB.108.054301, PhysRevB.95.094302, zhang2023information, PhysRevB.99.224307, PhysRevA.108.022415}. Most of the studies on many-body entanglement dynamics has been focusing on condensed matter systems, such as spin chains \cite{PhysRevA.66.032110,PhysRevA.70.062304, PhysRevA.69.022304, PhysRevA.69.034304, PhysRevB.79.104428, PhysRevA.64.012313, PhysRevA.84.022314, PhysRevA.97.042330} and all-to-all systems \cite{PhysRevB.108.054301, zhang2023information}, with others focusing on collision models \cite{PhysRevA.101.042324, Li_2022}, repeatedly measured random unitary circuits \cite{PhysRevB.99.224307, PhysRevX.9.031009}, etc. However, these studies usually share some common limitations. First, there is no reliable entanglement measure for bipartite and multipartite systems in  general mixed state cases. As a result, the entanglement measure used in literatures varies and each may have its own defects. For instance, many use entanglement entropy to characterize the entanglement between bipartite systems. The entanglement entropy is a reliable entanglement measure only when the total system of the bipartite systems is a pure state, restricting the scope of the study: One may want to study the bipartite entanglement within a subsystem of a larger pure system, but the subsystem in general will be a mixed state, hence entanglement entropy is not applicable. Some others use entanglement negativity as an entanglement measure \cite{PhysRevA.108.022415}, which is easy to compute, but it only serves as a sufficient, but not necessary, condition of entanglement. Some more recent works use bipartite mutual information and tripartite mutual information to study quantum information scrambling \cite{PhysRevA.101.042324,Li_2022, PhysRevA.97.042330,zhang2023information}, a process where local information, including entanglement, propagates into the environment. However, mutual information captures both classical correlations and quantum correlations, hence it is not a reliable entanglement measure, either. Second, many studies focus on the macroscopic behavior of entanglement dynamics, such as the growth of entanglement entropy with the size of the boundaries between two partitions \cite{RevModPhys.82.277}. In comparison, the microscopic behavior of entanglement dynamics has been largely overlooked. Moreover, in studies that do have a focus on the microscopic entanglement dynamics \cite{PhysRevA.70.062304,PhysRevA.66.032110,PhysRevA.69.034304,PhysRevA.69.022304,PhysRevB.79.104428,PhysRevA.64.012313,PhysRevA.84.022314, PhysRevA.89.022303}, the cause-and-effect relationships of entanglement generation, propagation and decay are often obscured by the many degrees of freedom present in condensed matter systems.


On the contrary, studies on few-body systems often better captures the microscopic behaviors of entanglement dynamics, such as the well-known entanglement sudden death \cite{PhysRevLett.93.140404} and entanglement genesis \cite{PhysRevA.78.062322}. Nevertheless, these studies typically share some common limitations, too. First, except for a few studies which focus solely on interactions among several qubits \cite{PhysRevA.68.052306, Yonac_2007, Gennaro_2008}, most studies introduce an environment, such as a thermal bath, that interacts with the few-body system and changes its entanglement usually in an undesirable way. The environment, being regarded as a noise, is usually traced out in the final calculations, meaning that the physical details about it are discarded. While this practice greatly reduces the technical difficulties of the study, one may be curious about what happens to the environment: Will the environment exhibit any form of entanglement, within itself or with the few-body system? If so, what factors affect the formation of the entanglement and how can we study its evolution through a microscopic lens? Answering these questions provides deeper insights into the nature of entanglement beyond the few-body system itself. Second, Wootter's concurrence \cite{PhysRevLett.80.2245} is commonly used in these studies as a reliable entanglement measure for the two-qubit pairwise entanglement. However, because of the absence of the generalization of the concurrence in more general cases, it is usually hard to apply the formalisms and techniques used in few-body systems to many-body systems, even numerically. On the other hand, it is unclear to what extent the findings in the few-body systems remain meaningful and applicable for many-body systems with more complicated interactions.
\subsection{\label{subsec:level2}Summary of main results}


In this article, we overcome the aforementioned common shortcomings. We study the entanglement dynamics of a family of quantum collision models \cite{CICCARELLO20221}. Collision models are special in that we can apply few-body physics techniques to explore the physics of certain many-body systems. Moreover, in a typical collision model, only one collision occurs at a time, meaning that we can monitor how a single collision affects the entanglement of the whole many-body system. This allows us to obtain a frame-by-frame, microscopic description of entanglement dynamics, enabling us to identify exactly what causes specific changes in the entangled states. This detailed insight into the dynamics of quantum entanglement sheds light on the understanding the fundamental processes in quantum systems, such as decoherence and information scrambling, at a microscopic level. It is worth noting it has been shown that collision models can efficiently simulate any multipartite Markovian quantum dynamics \cite{PhysRevLett.126.130403}, meaning that our results may readily describe the entanglement dynamics in various multipartite Markovian quantum dynamics. Furthermore, we develop a diagrammatic method, which provides a simple, intuitive and very general qualitative description on the pairwise entanglement dynamics scalable to many-body system including qudits.

On the other hand, in the multiple examples we will be showing, we identity a type of genuine multipartite entanglement, which we refer to as \textit{entanglement quilts}, where every qubit is entangled with every other qubit. Hence for a condensed matter system that is an entanglement quilt, it also shows long-range entanglement, which is related to quantum phase transition \cite{PhysRevA.66.032110}. It is found that entanglement quilts can be generated very simply in collision models, and the number of operations required scales linearly with the number of qubits.

The paper is organized as follows. In Section \ref{sec:level2}, we introduce the general formalism we will be using to study the collision models. The key idea is to utilize the invariant structures of certain density matrices in certain cases such that their entanglement is always easy to calculate, even after time evolutions.

In Section \ref{sec:level3}, we present multiple examples. Section \ref{subsec:level3} studies the entanglement dynamics of an excited qubit (system qubit) interacting with a group of ground state qubits (bath qubits) via the excitation-exchange type of Hamiltonian, $H_{ee}=\sigma_- \otimes \sigma_+ + \sigma_+ \otimes \sigma_-$. Although we refer to them as the system qubit and bath qubits for the reader's convenience, they are actually treated with equal consideration. We consider varied collision schemes, including system-bath interactions, bath-bath interactions and random interactions. The model can be considered as a process where the excited qubit is cooled down by a cold bath, and we show that regardless of the collision schemes, the whole system will become an entanglement quilt. Any new qubit colliding with any qubit within this entanglement quilt will also ``join in'' and become entangled with every other qubit. Some designed collision schemes provide a procedure to prepare W states with a linear scaling in complexity, easily implementable by a quantum computer.
Section \ref{subsec:level4} studies a similar model, but the bath contains one or more excited qubits. This modification is used to model the scenarios where the bath is not fully cooled down and contains thermal noise. Remarkably, it is found that the presence of excited bath qubits, even a single one, significantly alters the entanglement dynamics and largely destroys the entanglement quilt. The greater the proportion of excited bath qubits, the more fragmented and more localized the pairwise entanglement becomes. We provide a detailed analysis on this hypersensitivity behavior, which brings insight to the disappearance of long-range entanglement in many-body systems with a low but non-zero temperature \cite{PhysRevX.12.021022}. We estimate that for qubits with frequency $\sim$5 GHz, on average $\sim$$10^5$ qubits can be prepared to be as an entanglement quilt at $\sim$20 mK, which increases to $\sim$$10^{10}$ qubits at $\sim$10 mK. Section \ref{subsec:level5} studies how another type of interaction Hamiltonian, the spin-spin couplings $\sigma_x \otimes \sigma_x$, affect the entanglement dynamics. It is shown that the entanglement generated by this type of Hamiltonian tends to be highly localized because, different from the excitation exchange type of Hamiltonian $H_{ee}$, the total excitations are no longer conserved under this Hamiltonian, leaving the structure of the reduced density matrix a form harder to form entanglement. Section \ref{subsec:level6} extends the first example in Section \ref{subsec:level3} by allowing multiple qubits to interact with the same qubits simultaneously, namely many-to-one collisions, instead of one-to-one collisions. This model also produces entanglement quilts and we show analytically how the pairwise entanglement evolves after each collision.

In Section \ref{sec:level4}, we discuss multiple aspects on the formalism and the model. In Section \ref{subsec:level06}, we discuss the possible generalization to allow superimposed bath qubits. In Section \ref{subsec:level7}, we discuss and present an example where interactions between qubits, which are already entangled with other qubits, are allowed. By allowing interactions between correlated qubits, one can study the entanglement dynamics of the thermalization process where a system qubit interacts with a small bath. In Section \ref{subsec:level10}, we briefly discuss how our results can provide insights into condensed matter systems. In Section \ref{subsec:level9}, we show that although all W-like states are entanglement quilts, the converse is not true: An entanglement quilt does not have to be a W-like state. This finding may help us understand and preserve long-range entanglement in many-body systems. In Section \ref{subsec:level8}, we discuss a speculated property of entanglement quilts: Every subsystem of it is entangled with every other subsystem.

Finally, in Section \ref{sec:level5} we conclude our paper.

\section{\label{sec:level2}Formalism}
To start with, we use Wootter's concurrence \cite{PhysRevLett.80.2245} $C_{ij} \in [0,1]$ to measure the entanglement between the qubit pair $(i,j)$. Concurrence is a reliable monotonic measure of two-qubit entanglement, which indicates disentanglement for qubit pair $(i,j)$ if $C_{ij}=0$, and maximal entanglement if $C_{ij}=1$. It is worth noting that all bipartite entanglement measures on two-qubit system are actually equivalent, meaning that they will all give the same conclusion on whether a two-qubit system is more or less entangled than another two-qubit system. This is because for all pure two-qubit system, they can always be expressed as $\ket{\psi}=\cos{\theta} \ket{00}+\sin{\theta} \ket{11}$ by Schmidt decomposition, where $\theta \in [0, \pi/4]$, such that there is only one free parameter.

For a two-qubit system, its concurrence is given by \cite{PhysRevLett.80.2245}
\begin{equation} \label{eq:eq1}
	C=\max\{0,\sqrt{\lambda_1}-\sqrt{\lambda_2}-\sqrt{\lambda_3}-\sqrt{\lambda_4}\},\end{equation}
where $\lambda_1 \geq \lambda_2 \geq \lambda_3 \geq \lambda_4 \geq 0$ are the eigenvalues of $\rho \tilde{\rho}$. Here $\rho$ is the reduced density matrix of the qubit pair, and $\tilde{\rho} \equiv (\sigma_y \otimes \sigma_y) \rho^*(\sigma_y \otimes \sigma_y)$, where $\sigma_y$ is the Pauli-$y$ matrix, and $\rho^*$ denotes complex conjugate of $\rho$ in the $z$ basis.

As calculating concurrence involves solving for the eigenvalues of a $4 \times 4$ matrix, neat analytical results exist only in limited cases. Hence, to understand entanglement dynamics in an analytic manner scalable to a large number of qubits, we need to focus on the cases where the reduced density matrix $\rho_{ij}$ of any qubit pair $(i,j)$ has a structure that enables an easy calculation of concurrence. This brings our attention to the density matrices with the following structures, which we call the the ``X''\cite{yu2005evolution}, ``$\square$'', ``$\phi$'', and ``Q'' state, as derived from their appearances,
\begin{equation}
\begin{aligned} \label{eq:eq2}
		&\rho_X=\begin{pmatrix}
			\rho_{11} &0&0&\rho_{14}\\
			0 &\rho_{22}&\rho_{23}&0\\
			0&\rho_{32}&\rho_{33}&0\\
			\rho_{41}&0&0&\rho_{44}
		\end{pmatrix}, \quad
		\rho_\square=\begin{pmatrix}
			0 &0&0&0\\
			0 &\rho_{22}&\rho_{23}&\rho_{33}\\
			0&\rho_{32}&\rho_{33}&\rho_{34}\\
			0&\rho_{42}&\rho_{43}&\rho_{44}
		\end{pmatrix},
		\\
		&\rho_\phi=\begin{pmatrix}
			\rho_{11} &0&0&0\\
			0 &\rho_{22}&\rho_{23}&0\\
			0&\rho_{32}&\rho_{33}&0\\
			0&0&0&\rho_{44}
		\end{pmatrix}, \quad \rho_Q=\begin{pmatrix}
			0 &0&0&0\\
			0 &\rho_{22}&\rho_{23}&0\\
			0&\rho_{32}&\rho_{33}&0\\
			0&0&0&\rho_{44}
		\end{pmatrix}.
		\end{aligned}
\end{equation}
Note that the Q-state can be considered as a special $\phi$-state, which is a special X-state.
The concurrences of these density matrices are particularly easy to calculate:
\begin{equation} \label{eq:eq3}
\begin{aligned}
	C_X&=2\max \{0,|\rho_{23}|-\sqrt{\rho_{11}\rho_{44}}, |\rho_{14}|-\sqrt{\rho_{22}\rho_{33}}\}, \text{\cite{yu2005evolution}}\\
	C_\square&=2\max \{|\rho_{23}|, \sqrt{\rho_{22}\rho_{33}}\},\\
	C_\phi&=2\max \{0,|\rho_{23}|-\sqrt{\rho_{11}\rho_{44}}\},\\
	C_Q&=2|\rho_{23}|.
	\end{aligned}
\end{equation}
However, just because the concurrence is easy to calculate at one moment does not imply it will be easy to calculate at the next moment. Therefore, to analytically investigate the entanglement dynamics of a physical process at arbitrary time, we need to further require that the structure of the reduced density matrix $\rho$ remain unchanged after time evolutions. This requirement restricts the kinds of dynamics we may study, including the possible interaction Hamiltonians, system Hamiltonians, and the initial states prior to interactions.

As such, we will focus on the quantum collision model \cite{CICCARELLO20221} for the rest of the paper. In the simplest collision model, only two qubits interact with each other at a time and all of the rest qubits remain isolated so they do not interact with any qubit. After the initial round of collision, subsequent rounds proceed similarly, with potentially different pairs of qubits colliding with each other, one at a time. This model has gained increasing interests in recent years because of its simplicity and effectiveness in simulating and describing open quantum system dynamics. In particular, it has been shown that collision models can efficiently simulate any multipartite Markovian quantum dynamics \cite{PhysRevLett.126.130403}. Another reason is that quantum collision model focuses on sequential two-qubit interactions, much in resembling that of a quantum computer, wherein a set of two-qubit gates are sufficient to perform universal quantum computation.

\begin{figure}[t!]
\begin{center}
\begin{tikzpicture}
[x=0.75pt,y=0.75pt,yscale=-1,xscale=1]
\node (A) at (60,-50) {$B$};
\node (C) at (60,-00) {$C$};
\node (B1) at (0,-100) {$A_1$};
\node (B11) at (0,-130) {$$};
\node (B12) at (-32,-110) {$$};
\node (B13) at (-20,-125) {$$};
\node (B2) at (40,-100) {$A_2$};
\node (B21) at (40,-130) {$$};
\node (B22) at (60,-125) {$$};
\node (B23) at (20,-125) {$$};
\node (dots1) at (80,-100) {$\dots$};
\node (dots2) at (75,-75) {$\dots$};
\node (Bn) at (120,-100) {$A_n$};
\node (Bn1) at (120,-130) {$$};
\node (Bn2) at (100,-125) {$$};
\node (Bn3) at (140,-125) {$$};
\draw[decorate,decoration={coil,aspect=0,segment length=4pt,amplitude=0.6pt}] (A) --(B1);
\draw[decorate,decoration={coil,aspect=0,segment length=4pt,amplitude=0.6pt}] (A) --(B2);
\draw[decorate,decoration={coil,aspect=0,segment length=4pt,amplitude=0.6pt}] (A) --(Bn);
\draw (B11) --(B1);
\draw (B12) --(B1);
\draw (B13) --(B1);
\draw (B21) --(B2);
\draw (B22) --(B2);
\draw (B23) --(B2);
\draw (Bn1) --(Bn);
\draw (Bn2) --(Bn);
\draw (Bn3) --(Bn);
\draw [<-](A) edge (C);
\draw[dashed] (C) to[out=-150,in=90] (B1);
\draw[dashed] (C) to [out=-120,in=95](B2);
\draw[dashed] (C) to [out=-30,in=90] (Bn);
\end{tikzpicture}
\end{center}
\caption{\label{fig:epsart2}Diagrammatic description of entanglement dynamics in collision models where each letter denotes a qubit. The arrow  between B and C denotes they are interacting with each other and are potentially forming a new entanglement. The dashed lines between C and A$_i$ denote they used to be disentangled, but \textit{may} now become entangled because of the B--C interaction. The wavy lines between B and A$_i$ denote they used to be entangled, but their entanglement may have changed because of the B--C interaction. The straight lines connected to A$_i$ denotes the entanglement between A$_i$ and other particles (not shown), which are unaffected despite of the B--C interaction. Note that the figure is a very general qualitative description and also applies to qudits.}
\end{figure}
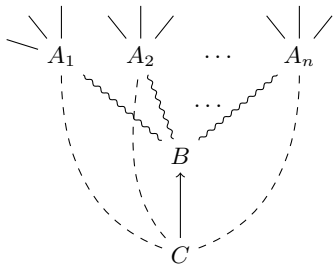

The focus on one interaction at a time in collision models means that to calculate pairwise concurrence, in the best cases, we only need to deal with at most a $2^3$--dimensional Hilbert space formed by three qubits each time. To understand this, in Fig.\,\ref{fig:epsart2} we provide a diagrammatic description. Each time a new qubit (the qubit C), which is not entangled with any other qubits, interacts with an old qubit (the qubit B). The old qubit used to have some interaction with other qubits so it is entangled with some qubits (the qubit A$_i$). Because of the B--C interaction, new entanglement between B--C and A$_i$--C may form, and old entanglement between A$_i$--B may change. Note how the entanglement between A$_i$--C may form nonlocally, even though they have never interacted with each other before. This phenomena has been reported in the few-qubit case \cite{Yonac_2007}.

The key to track the entanglement dynamics here is to obtain a recurrence relation between the pre-collision reduced density matrix $\rho$ and the post-collision reduced density matrix $\rho^\prime$.
For instance, to calculate the post-collision entanglement between B--C, A$_i$--B and A$_i$--C,
 we need to consider the three-qubit density matrix $\rho_{A_iBC}$ and calculate $\rho^\prime_{A_iBC}=e^{-iH_{A_iBC}t} \rho_{A_iBC}e^{iH_{A_iBC}t}$, where $H_{A_iBC}$ is total time-independent Hamiltonian of A$_i$, B and C. After obtaining $\rho^\prime_{A_i BC}$, we perform a partial tracing to obtain $\rho^\prime_{A_i B}$, $\rho^\prime_{A_i C}$ and $\rho^\prime_{BC}$, with which we can calculate the pairwise concurrence. Note that the entanglement between A$_i$ and A$_j$ remain unchanged, because their total Hamiltonian $H_{A_i A_j}=H_{A_i}+H_{A_j}$ can be factored into commuting local Hamiltonians in their subspace, and local Hamiltonian does not change the entanglement.

Recall that to calculate concurrence analytically, we rely on the special structures [Eq.\,(\ref{eq:eq2})] of the reduced density matrices, and such structures must remain unchanged after interactions. This requirement leads us to study the following interaction Hamiltonians, $H_{xy}$ and $H_{ee}$ (``ee'' stands for excitation exchange), as a special case of the two-qubit dynamics,
\begin{equation} \label{eq:eq4}
\begin{aligned}
	H_{xy}&=\sqrt{\Omega}(\cos{\theta}\sigma_{B,x}+\sin{\theta}\sigma_{B,y})\otimes(\cos{\theta}\sigma_{C,x}+\sin{\theta}\sigma_{C,y})\\
	H_{ee}&=\Omega (\sigma_{B,+} \otimes \sigma_{C,-}+\sigma_{B,-} \otimes \sigma_{C,+})\\
	&=\Omega(\sigma_{B,x} \otimes \sigma_{C,x}+\sigma_{B,y} \otimes \sigma_{C,y}),
\end{aligned}
\end{equation}
where $\Omega \in \mathbb{R^+}$ is the interaction strength (with an appropriate unit), $\theta$ is a real parameter, and $\sigma_+=\ket{1}\bra{0}$ and $\sigma_-=\ket{0}\bra{1}$. As for the qubits' own Hamiltonians, we may consider the inhomogeneous case, where each qubit has its own frequency $\omega_i$, such that for the $i$th qubit its system Hamiltonian is given by $H_i=\frac{1}{2}\omega_i \sigma_{C_i,z}$, or the homogenous case, where all qubits share the same frequency $\omega$.

Finally, to satisfy the requirement on the invariant structures of the density matrix, we require the initial states of the qubits to be $\ket{0}$ or $\ket{1}$. In Section \ref{subsec:level06} we discuss what if the initial states of the qubits are superimposed, i.e. $e^{i\phi} \cos\theta \ket{0}+\sin \theta \ket{1}$. In Table \ref{tab:table1}, we summarize the types of dynamics one may study analytically, with their recurrence relations shown in detail in the Table \ref{tab:appendixtable3}, \ref{tab:appendixtable2}, and \ref{tab:appendixtable4} in Appendix \ref{sec:appendix1}-\ref{sec:appendix3}. Note that unless otherwise specified, for each collision we only allow the new qubit, which is disentangled with any other qubits, to interacts with the old qubit. This is because the interaction dynamics between old qubits can be much more complicated; see Section \ref{subsec:level7}.

\begin{table}[!t]
  \centering
  \caption{Some of the known types of pairwise interaction which keep the structure of the density matrix unchanged. Other examples include $H_{xy}+H_{zz}$, which keeps the X-structure invariant, and $H_{ee}+H_{zz}$, which keeps all four structures invariant. Here $H_{zz}=\sigma_{B,z} \otimes \sigma_{C,z}$.}
    \renewcommand{\arraystretch}{1.1}
    \begin{tabular}{cccccc}
    \toprule
    No.&Str. & $H_\text{int}$     & $H_\text{sys}$ & New qubit &Ref. Tab.\\
    \midrule
    1&$\rho_X$     &  $H_{xy}$     &     \multirow{5}{*}{\shortstack{homo./\\inhomo.}}  &  $\ket{0}$, $\ket{1}$ &\ref{tab:appendixtable3}\\
   2&$\rho_X$     &   $H_{ee}$    &       & $\ket{0}$, $\ket{1}$ & \,\ref{tab:appendixtable2}\\ 
    3&$\rho_\phi$     &  $H_{ee}$     &       &$\ket{0}$, $\ket{1}$  & \,\ref{tab:appendixtable2}\\
    4&$\rho_Q$     &  $H_{ee}$     &      &  $\ket{0}$ only &\,\ref{tab:appendixtable2}\\
    5&$\rho_{\square}$     &    $H_{ee}$   &       & $\ket{0}$ only &\,\ref{tab:appendixtable4}\\
    \bottomrule
    \end{tabular}%
  \label{tab:table1}%
\end{table}%

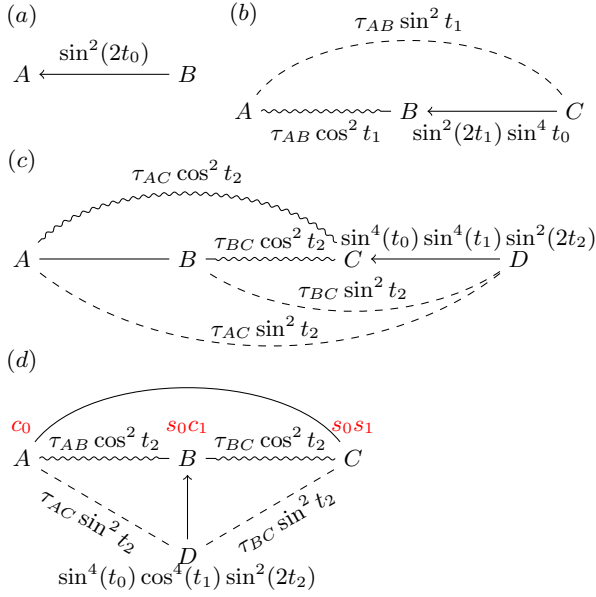
\begin{figure}[!t]
\centering
\begin{adjustbox}{width=0.45\textwidth}	
    \begin{tikzpicture}[x=0.75pt,y=0.75pt,yscale=-1,xscale=1]
        \begin{scope}[shift={(2cm, 2cm)}] 
            \begin{scope}[xshift=0cm, yshift=-0.5cm]
                \node (a) at (0,-30) {$(a)$};
				\node (A) at (0,0) {$A$};
				\node (B) at (85,0) {$B$};
				\draw  [<-]  (A) edge (B) ;
				\node[above] at ($(A)!0.5!(B)$) {$\sin^2 (2t_0)$};
            \end{scope}

            \begin{scope}[xshift=3cm, yshift=0cm]
               \node (b) at (0,-50) {$(b)$};
				\node (A) at (0,0) {$A$};
				\node (B) at (85,0) {$B$};
				\node (C) at (170,0) {$C$};
				\node (eq1) at (85,-47) {$\tau_{AB}\sin^2 t_1 $};
				\draw[->] (C) edge (B);
				\draw[dashed]  (C) to[out=-120,in=-60] (A);
				\draw [decorate,decoration={coil,aspect=0,segment length=4pt,amplitude=0.6pt}]   (A) -- (B);
				\node[below] at ($(B)!0.5!(C)$) {$ \sin ^{2}(2t_1)\sin^4t_0$};
				\node[below] at ($(A)!0.5!(B)$) {$\tau_{AB}\cos^2 t_1 $};
            \end{scope}

            \begin{scope}[xshift=0cm, yshift=2cm]
                \node (c) at (0,-50) {$(c)$};
				\node (A) at (0,0) {$A$};
				\node (B) at (85,0) {$B$};
				\node (C) at (170,0) {$C$};
				\node (D) at (255,0) {$D$};
				\node (eq1) at (85,-45) {$\tau_{AC}\cos^2 t_2 $};
				\node (eq2) at (170,15) {$\tau_{BC}\sin^2 t_2 $};
				\node (eq3) at (125,35) {$\tau_{AC}\sin^2 t_2 $};
				\draw[decorate,decoration={coil,aspect=0,segment length=4pt,amplitude=0.6pt}] (C) --(B);
				\draw[->] (D) edge (C);
				\draw[decorate,decoration={coil,aspect=0,segment length=4pt,amplitude=0.6pt}]  (C) to[out=-125,in=-55] (A);
				\draw[dashed]  (D) to[out=135,in=45] (B);
				\draw[dashed]  (D) to[out=135,in=45] (A);
				\draw    (A) -- (B);
				\node[above] at ($(B)!0.5!(C)$) {$\tau_{BC}\cos^2 t_2 $};
				\node[above] at ($(C)!0.7!(D)$) {$ \sin^4 (t_0)\sin^4 (t_1)\sin^2(2 t_2)$};
            \end{scope}

            \begin{scope}[xshift=0cm, yshift=4.7cm]

               \node (d) at (0,-50) {$(d)$};
				\node (A)[label=above:\textcolor{red}{$ c_0$}] at (0,0) {$A$};
				\node (B)[label=above:\textcolor{red}{$  s_0c_1$}] at (85,0) {$B$};
				\node (C)[label=above:\textcolor{red}{$s_0 s_1$}] at (170,0) {$C$};
				\node (D) at (85,50) {$D$};
				\node[rotate=30] (eq2) at (135,31) {$\tau_{BC}\sin^2 t_2 $};
				\node[rotate=-30] (eq3) at (35,31) {$\tau_{AC}\sin^2 t_2 $};
				\draw[->] (D) edge (B);
				\draw  (C) to[out=-120,in=-60] (A);
				\draw[dashed]  (D)--(A);
				\draw[dashed]  (D)--(C);
				\draw[decorate,decoration={coil,aspect=0,segment length=4pt,amplitude=0.6pt}]    (A) -- (B);
				\draw[decorate,decoration={coil,aspect=0,segment length=4pt,amplitude=0.6pt}] (C) --(B);
				\node[above] at ($(B)!0.5!(C)$) {$\tau_{BC}\cos^2 t_2 $};
				\node[above] at ($(A)!0.5!(B)$) {$\tau_{AB}\cos^2 t_2 $};
				\node[below] at ($(B)!1!(D)$) {$ \sin^4 (t_0)\cos^4 (t_1)\sin^2(2 t_2)$};
            \end{scope}
        \end{scope}
    \end{tikzpicture}
    \end{adjustbox}
    \caption{\label{fig:epsart3}The entanglement dynamics of the 4th collision model in Table \ref{tab:table1}, assuming all qubits have the same frequency $\omega$ and collides with each other via the excitation exchange type of Hamiltonian with the interaction strength $\Omega=1$. Initially, qubit A is at $\ket{1}$ and all other qubits are at $\ket{0}$. Panel (a): A--B interaction; (b): A--B interaction, followed by B--C interaction (abbreviated as ``\{A--B, B--C\} interaction''); (c): \{A--B, B--C, C--D\} interaction; (d): \{A--B, B--C, B--D\} interaction. The formulae along the lines denote the post-collision entanglement, as measured by tangle (the square of concurrence) $\tau=C^2$. For instance, in Panel (b), the post-collision entanglement $\tau^\prime_{AC}$ of A--C is given by $\tau^\prime_{AC}=\tau_{AB} \sin^2 t_1$, where $\tau_{AB}$ is the pre-collision entanglement between A--B, and $t_1$ is the time duration of the B--C collision. The red texts in Panel (d) is a shorthand to track the mean excitation of each qubit; see the main text.}
\end{figure}

\section{\label{sec:level3}Example}
\subsection{\label{subsec:level3}Qubit cooled down by a cold bath}
To demonstrate our formalism, in the following we show an example where the entanglement dynamics enjoys a set of particularly simple recurrence relations and produces \textit{entanglement quilts}. Consider the 4th model in Table \ref{tab:table1}, where the qubit-qubit interaction is given by the excitation exchange type of Hamiltonian, $H_{ee}$. Assume the qubits' own Hamiltonians are homogenous, such that all qubits share the same frequency $\omega$. Let the qubit A be prepared at $\ket{1}$, and all other qubits, denoted as B, C, etc., be at $\ket{0}.$ In this case, each collision in the model is isomorphic to the vacuum Jaynes-Cummings model under rotating wave approximation with vanishing detuning frequency, and the ground-state qubits can be regarded as the vacuum state of the photon.

In Fig.\,\ref{fig:epsart3}, we show diagrammatically the pairwise entanglement dynamics of this model after a series of collisions. Note how the new entanglement generates and old entanglement decays exactly in the way as described in Fig.\,\ref{fig:epsart2}. Here, we use tangle (the square of concurrence) $\tau=C^2$ to measure the entanglement, because in this special example, $\tau$ is a conserved quantity among nonlocal change of entanglement, i.e. $\tau^\prime_{A_iC}+\tau^\prime_{A_iB}=\tau_{A_iB}$ [see Eq.\,(\ref{eq:eq5}) and (\ref{eq:eq05})], where $\tau$ and $\tau^\prime$ denote pre- and post-collision tangle, respectively. In general, the entanglement dynamics in this model follows a set of simple relations, as exemplified by Fig.\,\ref{fig:epsart2}:
\begin{equation} \label{eq:eq5}
\begin{aligned}
	\tau^\prime_{BC}&=4\rho^2_{33} \sin^2(\Omega_{BC}t)\cos^2(\Omega_{BC}t)\\
	\tau^\prime_{A_iC}&=\tau_{A_iB} \sin^2(\Omega_{BC}t)\\
	\tau^\prime_{A_iB}&=\tau_{A_iB} \cos^2( \Omega_{BC}t),
	\end{aligned}
\end{equation}
where $t$ is the time duration of B--C interaction, $\rho_{33}$ is the entry (3,3) of pre-collision density matrix $\rho_{AB}$, and it is equal to the mean excitation of qubit B, i.e. $\bra{1} \rho_B\ket{1}=\rho_{33}$. 

As $\rho_{33}$ appears in the above dynamical relations, one needs to track the mean excitation of each qubit. Here we provide a shorthand for easy tracking. For each collision, assign a $c_i$ (stands for $\cos^2(t_i)$) to the colliding odd qubit and a $s_i$ (stands for $\sin^2(t_i)$) to the new qubit. The accumulated assigned symbols then represent the mean excitation. For instance, in Fig.\,\ref{fig:epsart3}(d), qubit B owns $s_0c_1$, therefore $\bra{1}\rho_B\ket{1}=\sin^2(t_0)\cos^2(t_1) $ before B--D interaction.

In the above, we have considered the case where all qubits share the same frequency $\omega$. One can also obtain the recurrence relations of the tangle when this is not true:
\begin{equation} \label{eq:eq05}
\begin{aligned}
	\tau^\prime_{BC}&=4\rho^2_{33} \beta_s(t) \beta_c(t)\\
	\tau^\prime_{A_iC}&=\tau_{A_iB} \beta_s(t)\\
	\tau^\prime_{A_iB}&=\tau_{A_iB} \beta_c(t),
	\end{aligned}
\end{equation}
where
\begin{equation}
	\begin{aligned}
		\beta_s &=\frac{\Omega_{BC}^2 \sin^2(t \Delta)}{\Delta^2} \\
		\beta_c &=1-\beta_s \\
		\Delta &=\sqrt{\frac{1}{4}\left(\omega_{B}-\omega_{C}\right)^{2}+\Omega_{BC}^{2}}.
	\end{aligned}
\end{equation}
In general, a nonzero detuning frequency $\delta=\omega_B-\omega_C$ makes it harder for the old entanglement between A$_i$--B to decay and new entanglement between A$_i$--C and B--C to form. Moreover, it makes it impossible for the old entanglement between A$_i$--B to completely disappear for any time $t$.

\begin{figure*}[!tbp]
\includegraphics[width=0.99\textwidth]{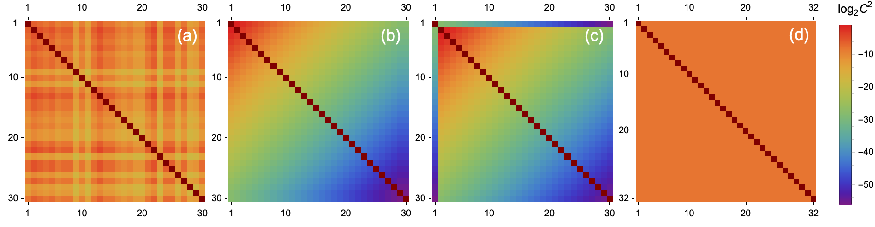}
\caption{
\label{fig:epsart4} Visualization of pairwise entanglement of the 4th model in Table \ref{tab:table1} as measured by $\log_2{C^2}$ (log of concurrence squared) of 30 qubits after 29 collisions (32 qubits after 31 collisions for panel (d)). Each pixel represents the pairwise entanglement between two qubits. For instance, pixel (2,3) represents the entanglement between the 2nd and 3rd qubit. The dark red pixels denote disentanglement. Panel (a): each new qubit collides with a random old qubit; panel (b): the $i$th qubit collides with the $(i-1)$-th qubit; panel (c): the $i$th qubit collides with the 1st qubit; panel (d): the new qubits collides with the \{1,1,2,1,2,3,4,1,2,3,4,5,6,7,8,\dots\}-th qubit in sequence. The same result of panel (d) can also be achieved by letting the $i$th qubit collides with the $(i-1)$-th qubit with collision time $t=\arcsin \sqrt{\frac{n-i+1}{n-i+2}}$ where $n$ is the total number of qubits. Note how there creates an entanglement quilt after a series of collisions: Every qubit becomes entangled with every other qubits. In particular, panel (d) exhibits uniform entanglement quilt: Every qubit is equally entangled with every other qubits. Here we take the interaction time $t=\pi/4$ unless otherwise specified, interaction strength $\Omega=1$, qubit frequency $\omega=1$. Initial state: the 1st qubit at the excited state, and all other qubits at the ground states.
 } 
\end{figure*}

As the entanglement dynamics of this model enjoys a set of simple recurrence relations, one can easily calculate the pairwise entanglement for all qubit pairs after many collisions.  In Fig.\,\ref{fig:epsart4}, we show the pairwise entanglement of this model after multiple collisions via heat maps under varied colliding rules. The pixel $(i,j)$ of each heat map presents the pairwise entanglement between the $i$th and $j$th qubits.

There are several comments worth mentioning regarding Fig.\,\ref{fig:epsart4}. First, it can be shown that regardless of the colliding rules and the number of collisions, in this particular model, the state $\ket{\psi}$ for the whole system is always a W-like state,
\begin{equation}
	\ket{\psi}=a_1\ket{100\dots0}+a_2\ket{010\dots0}+\dots+a_n\ket{000\dots1},
\end{equation}
where $\sum_{i=1}^n |a_i|^2=1.$ This suggests that the effective dimension of the Hilbert space is just $n$, where $n$ is total number of qubits. Second, we note that after a series of collisions, there creates an entanglement quilt: Every qubit becomes entangled with every other qubit. Moreover, an entanglement quilt is also a genuine multipartite entanglement, meaning that it is not $m$-separable for any $m > 1$. Note that while all W-like states with nonzero $a_i$ for all $i$ are entanglement quilts, the converse is not true --- an entanglement quilt does not have to be a W-like state; see Section \ref{subsec:level9} for a counterexample. Hence, the full properties of entanglement quilts are still to be explored; for discussions see Section \ref{subsec:level8}.

Third, Fig.\,\ref{fig:epsart3} and Fig.\,\ref{fig:epsart4} suggest procedures to generate W-like states (hence an entanglement quilt). Notably, in Fig.\,\ref{fig:epsart4}(d), by designing the order and time of collisions, one can generate an uniform entanglement quilt, where every qubit is equally entangled with every other qubits. This state turns out to be the W state up to a local phase, such that $a_j \in \{\frac{1}{\sqrt{n}}, \frac{i}{\sqrt{n}}, -\frac{1}{\sqrt{n}}, -\frac{i}{\sqrt{n}}\}$ for $j=1, 2,\dots, n$. The key to preparing the state is by letting each qubit have the same mean excitation, indicating that the amount of any two-qubit entanglement is small as $n$ grows, which is in accordance with the monogamy of entanglement [Eq.\,(\ref{eq:eq30})]. This could be performed, for example, by carefully controlling the collision time $t$ between the $i$th and $(i-1)$-th qubit, such that $t=\frac{1}{\Omega}\arcsin \sqrt{\frac{n-i+1}{n-i+2}}$. Alternatively, if the total number of qubits are $n=2^m$ for some integer $m$, one can let the new qubits successively collide with the $\{1,1,2,1,2,3,4,1,2,3,4,5,6,7,8,\dots\}$-th qubit for time $t=\pi/4\Omega$, with their two-qubit gate being \begin{equation} \label{eq:eq6}
	\exp\left(-i H_{ee} \frac{\pi}{4\Omega}\right)=\begin{pmatrix}
		1&0&0&0\\
		0&\frac{1}{\sqrt{2}}&-\frac{i}{\sqrt{2}}&0\\
		0&-\frac{i}{\sqrt{2}}&\frac{1}{\sqrt{2}}&0\\
		0&0&0&1
	\end{pmatrix},
\end{equation}
which equals to the converse of $\sqrt{\text{iSWAP}}$. In both cases, the number of operations required scale linearly with the number of qubits. 

Fourth, we note that Fig.\,\ref{fig:epsart3}(c), where all new qubits interact with the first qubit successively, can be interpreted as an excited qubit being cooled down by a cold bath consisting of an ensemble of ground-state qubits. The cold bath is large enough such that each bath qubit only have a chance to collide once with the first, excited qubit, with potentially many other bath qubits yet to be engaged in interactions.
(See Fig.\,\ref{fig:epsart12} for results of a small cold bath instead.) Different from typical studies of such cooling process where the main focus is the system qubit, here we pay equal attentions to both the system and bath qubits. In particular, we are curious what happens to the \textit{bath} qubits when they are used to cool down an excited qubit. It turns out that the cost the bath qubits pay during the cooling process is that they become an entanglement quilt, as well as being weakly entangled with the first qubit they are trying to cool down.

\begin{figure*}[!tbp]
\includegraphics[width=0.99\textwidth]{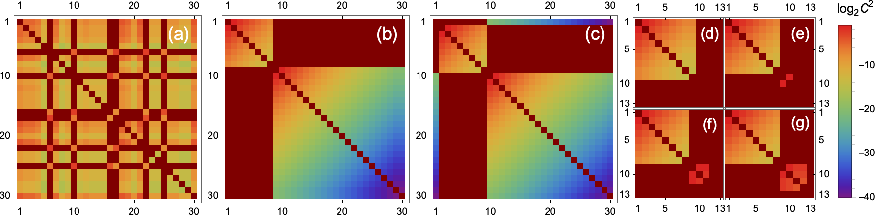}
\caption{
\label{fig:epsart5} Visualization of pairwise entanglement of the 3rd model in Table \ref{tab:table1} as measured by $\log_2{C^2}$. The dark red pixels denote disentanglement. Panels (a)-(c) have the similar setup as in Fig.\,\ref{fig:epsart4}(a)-(c), with the only difference being the 10th qubit here is initially at $\ket{1}$, instead of $\ket{0}$. Panels (d)-(g), which have been truncated, show snapshots of entanglement dynamics of panel (b) after the 8th--11th collisions. Note how the pairwise entanglement is fragmented into local blocks in the presence of a single excited qubit.} 
\end{figure*}

\begin{figure*}[!tbp]
\includegraphics[width=1\textwidth]{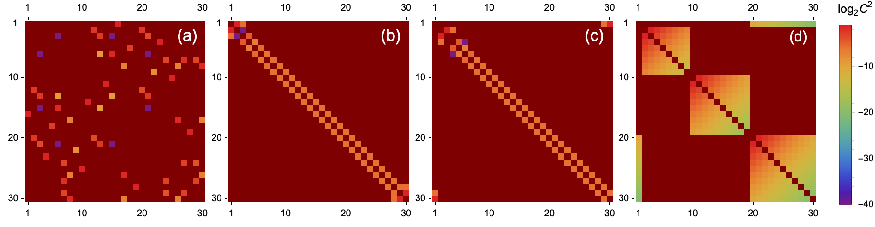}
\caption{
\label{fig:epsart6} Visualization of pairwise entanglement of the 3rd model in Table \ref{tab:table1} as measured by $\log_2{C^2}$. The dark red pixels denote disentanglement. Panels (a)-(c) have the similar setup as in Fig.\,\ref{fig:epsart4}(a)-(c), with the difference being that the qubits are initially prepared to be $\ket{10101\dots}_{ABCDE\dots}$. Panel (d) has the similar setup as (c), with the difference being that the 1st, 10th and 20th qubits are prepared to be $\ket{1}$ and all other qubits $\ket{0}$.} 
\end{figure*}

Finally, one can obtain analytical expressions of the pairwise entanglement after many collisions. For example, assuming the $i$-th qubit always collides with the $(i-1)$-th qubit [Fig.\,\ref{fig:epsart3}(b)], and we are interested in the entanglement $\tau_{1,m}$ between the 1st and the $m$th qubit after $m-1$ collisions. By making use of the recurrence relation $\tau_{1,m}=\tau_{1,m-1}\sin^2(\Omega_{m-1,n} t_{m-1})$, one obtains
\begin{equation} \label{eq:eq8}
	\tau_{1,m}=\sin^2(2\Omega t) \sin^{2(m-2)}(\Omega t),
\end{equation}
where interaction strength $\Omega$ and interaction time $t$ have been assumed to be equal for all collisions. The above result indicates that the nonlocal formation of entanglement (i.e. entanglement formed without direct interactions) decays exponentially with the number of collisions, if the collision time  is a constant. However, as has been pointed out in the previous paragraph, if the collision time is carefully controlled, then one can create a uniform entanglement quilt. In this case, the nonlocal formation of entanglement decays quadratically, instead of exponentially, with the number of qubits.


It is worth noting that the entanglement always decay asymptotically in this model, meaning that it takes infinite time for the tangle to decrease to zero. However, as we will see in the next example, the entanglement behavior of this model is hypersensitive to thermal fluctuations, and it displays entanglement sudden death in the presence of even a single excited bath qubit.


\begin{figure*}
\includegraphics[width=\linewidth]{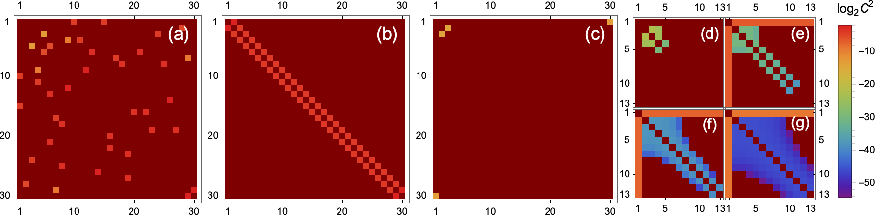}
\caption{
\label{fig:epsart7} Visualization of pairwise entanglement of the 1st model in Table \ref{tab:table1} as measured by $\log_2{C^2}$ of 30 qubits after 29 collisions. The dark red pixels denote disentanglement. Panels (a)-(c) have the similar collision schemes as in Fig.\,\ref{fig:epsart4}(a)-(c), but with the collision time $t=\pi/8$. Panels (d)-(g), which have been truncated, have the similar setup as panel (c), but with the collision time $t=\pi/16, \pi/32, \pi/64, \pi/128$, respectively. This model is insensitive to local fluctuations of excitations: If the initial state is prepared as the N\'eel state $\ket{10101\dots}_{ABCDE\dots}$, the results will look mostly the same. Note how panels (d)-(g) exhibit a sensitivity on the collision time: Shorter collisions times tend to generate and keep nonlocal entanglement better. However, this is exclusive to the collision scheme deployed in panel (c) --- if the same shorter collision times are deployed in panel (a) and (b), the results show no significant changes from the originals.} 
\end{figure*}

\subsection{\label{subsec:level4}Qubit cooled down with thermal fluctuations}
Inspired from the previous model, one may be tempted to think that the formation of entanglement quilt is ubiquitous in Nature. However, this is not true. On the contrary, the entanglement quilt is very sensitive and fragile to the local fluctuation of excitations of new qubits. Below, we provide an example followed by a detailed analysis to illustrate the sudden disappearance of the original long-range entanglement and the inability to form new long-range entanglement in the presence of a single excited qubit.

Let us consider the pairwise entanglement under the same dynamics and similar setup as in Fig.\,\ref{fig:epsart4}, with the only difference being that the 10th qubit is initially prepared to be at $\ket{1}$, rather than $\ket{0}$. This single modification alters the dynamics such that it belongs to the 3rd model, instead of the 4th,  in Table \ref{tab:table1}. In Fig.\,\ref{fig:epsart5}, we show the heat maps of the pairwise entanglement under this modification. The appearance of just a single excited qubit has largely destroyed the entanglement quilt. The entanglement among the qubits are broken into blocks, and the entanglement quilt survives only within the block (note that the qubits in Fig.\,\ref{fig:epsart5}(a) can be renumbered and rearranged to form two blocks), meaning that the pairwise entanglement becomes more localized.

The phenomenon can be explained mathematically as follows. Consider, for example, the collision rule specified in Fig.\,\ref{fig:epsart5}(b), where the $i$th qubit collides with the $(i-1)$-th qubit. After $m-1$ collisions but right before the collision with the $(m+1)$-th excited qubit (e.g. the 10th qubit in Fig.\,\ref{fig:epsart5}(b)), the reduced density matrix between the $m$th and $j$th ($j<m$) qubit is given by
\begin{equation} \label{eq:eq9}
	\begin{aligned}
		\tilde{\rho}_{22}:=[\rho_{jm}]_{22}&=\sin^{2(j-1)}(\Omega t) \cos^2(\Omega t)\\
		\tilde{\rho}_{33}:=[\rho_{jm}]_{33}&=\sin^{2(m-1)}(\Omega t)\\
		\tilde{\rho}_{44}:=[\rho_{jm}]_{44}&=1-[\rho_{jm}]_{22}-[\rho_{jm}]_{33}\\
		\tilde{\rho}_{23}:=[\rho_{jm}]_{23}&=[\rho_{jm}]^*_{32}=i^{n-j} \sin^{j+m-2}(\Omega t)\cos (\Omega t),
	\end{aligned}	
\end{equation}
and all other entries vanish. In the above we have assumed that the interaction strength $\Omega$ and interaction time $t$ remain the same for all collisions. Next, the $(m+1)$-th qubit, which is excited, collides with the $m$th qubit. Then, the post-collision reduced density matrix $\rho^\prime_{jm}$ and $\rho^\prime_{j,m+1}$ are given by
\begin{widetext}
\begin{equation} \label{eq:eq10}
	\rho^\prime_{jm}=\begin{pmatrix}
		\tilde{\rho}_{22} \sin^2(t \Omega)&0&0&0\\
		0&\tilde{\rho}_{22} \cos^2(t \Omega) &\tilde{\rho}_{23} \cos(t\Omega)&0\\
		0&\tilde{\rho}_{32} \cos(t\Omega)&\tilde{\rho}_{33}+\tilde{\rho}_{44} \sin^2(t \Omega) &0\\
		0&0&0&\tilde{\rho}_{44} \cos^2(t \Omega) 
	\end{pmatrix}
\end{equation}

\begin{equation} \label{eq:eq11}
	\rho^\prime_{j,m+1}=\begin{pmatrix}
		\tilde{\rho}_{22} \cos^2(t \Omega)&0&0&0\\
		0&\tilde{\rho}_{22} \sin^2(t \Omega) &-i\tilde{\rho}_{23} \sin(t\Omega)&0\\
		0&i\tilde{\rho}_{32} \sin(t\Omega)&\tilde{\rho}_{33}+\tilde{\rho}_{44} \cos^2(t \Omega) &0\\
		0&0&0&\tilde{\rho}_{44} \sin^2(t \Omega)
	\end{pmatrix}.
\end{equation}
\end{widetext}

A major change to the reduced density matrices is that because of the collision with the excited qubit, $\rho^\prime_{jm}$ and $\rho^\prime_{j,m+1}$ are no longer $Q$-states, but $\phi$-states. This results in a different expression of their concurrences. Compared with $C_Q$, $C_\phi$ has an extra $-\sqrt{\rho_{11}\rho_{44}}$ term [Eq.\,(\ref{eq:eq3})], which always contributes negatively to the concurrence. In other words, compared with $Q$-states, it is in general harder for $\phi$-states to be an entangled state. In fact, one can calculate $C^\prime_{jm}$ and $C^\prime_{j,m+1}$:
\begin{widetext}
\begin{equation} \label{eq:eq12}
\begin{aligned}
	C^\prime_{jm}&=2\max\{0|\tilde{\rho}_{23} \cos(t \Omega)|-\sqrt{\tilde{\rho}_{22}\tilde{\rho}_{44}}|\cos(t \Omega) \sin(t \Omega)| \} \\
	&\approx 2 \max\{0, |\cos^2 (t\Omega) \sin^{j+m-2} (\Omega t)|-|\cos^2(\Omega t) \sin^{j-1}(\Omega t)|\} \quad \text{for $m$ large}
	\end{aligned}
	\end{equation}
	\begin{equation} \label{eq:eq13}
	\begin{aligned}
	C^\prime_{j,m+1}&=2\max\{0, |\tilde{\rho}_{23} \sin(t \Omega)|-\sqrt{\tilde{\rho}_{22}\tilde{\rho}_{44}}|\cos(t \Omega) \sin(t \Omega)|\}\\
	&\approx 2 \max\{0, |\cos (t\Omega) \sin^{j+m-1} (\Omega t)|-|\cos^2(\Omega t) \sin^{j-1}(\Omega t)|\} \quad \text{for $m$ large}
	\end{aligned}
\end{equation}
\end{widetext}
When $m$ is large, $\tilde{\rho}_{33}$ is very small, and we can take $\tilde{\rho}_{44} \approx 1-\tilde{\rho}_{22}$, such that $\sqrt{\tilde{\rho}_{22}\tilde{\rho}_{44}} \approx |\cos(\Omega t)\sin^{j-1}(\Omega t)|$, where we have used Taylor expansion $\sqrt{(1-x)x}\approx \sqrt{x}$. This implies $|\tilde{\rho}_{23}| \ll \sqrt{\tilde{\rho}_{22}\tilde{\rho}_{44}}$, as is also evident in Eq.\,(\ref{eq:eq12}) and Eq.\,(\ref{eq:eq13}), meaning that we almost always have $C^\prime_{jm}=C^\prime_{j,m+1}=0$. In other words, not only does the original entanglement between the $j$th and $m$th qubit suddenly disappears, but also there is no new nonlocal entanglement forming. The only new entanglement formed is between the $m$th and $(m+1)$th qubit, due to their direct interactions.

The above analysis has been vividly reflected in Fig.\,\ref{fig:epsart5}(d)-(g), where we show the time-lapse snapshots of the heat maps of the pairwise entanglement between the 8th and 11th collisions. As can be seen from panel (d) and (e), which record the entanglement before and after the collision with the excited qubit, all old entanglements between $j$th qubit, where $j<9$, and 9th qubit, disappear. Moreover, there is no new entanglement forming between the 10th qubit and $j$th qubit. The 10th qubit is only entangled with the 9th qubit via their local interaction.

If we increase the number of excited qubits, assuming they are scattered, rather than staying as a cluster among the other qubits, then the number of blocks also increases accordingly. For instance, if both the 10th and the 20th qubits are initially excited, then there will be three blocks [Fig.\,\ref{fig:epsart6}(d)]. In the extreme case, where the ground state qubit and excited state qubit occurs alternatingly, i.e. the N\'eel state $\ket{10101\dots}_{ABCDE\dots}$, then almost all nonlocal entanglement is destroyed and only local ones can survive [Fig.\,\ref{fig:epsart6}(a)-(c)].

However, if the 2nd-10th qubits are initially prepared at $\ket{0}$, and all other qubits are initially prepared at $\ket{1}$, then the scenarios like Fig.\,\ref{fig:epsart6}(b) and Fig.\,\ref{fig:epsart6}(c) will not occur. Instead, its entanglement behavior will be the same as Fig.\,\ref{fig:epsart5}(b) and Fig.\,\ref{fig:epsart5}(c), assuming the same collision schemes. In other words, what is important here is the local fluctuations of excitations. To form entanglement nonlocally (i.e. without direct interactions), the qubits always favor a locally homogenous environment, where all their surrounding qubits are initially at the ground state (or excited state). This is because the Q-state and $\square$-state remain invariant under interactions with $\ket{0}$ via the excitation exchange type of Hamiltonian $H_{ee}$; similarly for the ``flipped'' Q-state and $\square$-state interacting with $\ket{1}$.

Since the entanglement quilt formation is very sensitive to local thermal fluctuations, when one is trying to prepare an entanglement quilt experimentally, it is important to cool down the bath qubits sufficiently. Suppose one's goal is to prepare the ground state bath qubits such that 1 in $n$ qubits is excited. Assuming the system Hamiltonian of the bath qubit is given by $H=\frac{1}{2}\omega \sigma_z$, then by making use of $\rho=\operatorname{diag}(1/n, 1-1/n)=\exp(-H/k_BT)/\operatorname{Tr}[\exp(-H/k_BT)]$, where $k_B$ is the Boltzmann constant and $T$ is the temperature, one obtains,
\begin{equation} \label{eq:eq14}
	T(n)=\frac{\omega}{k_B \log(n-1)},
\end{equation}
which logarithmically decreases as $n$ increases. For example, in superconducting qubits with a typical frequency $\sim$ 5 GHz, we have $T(10^2) \simeq 50 \text{mK}$, $T(10^5) \simeq 20 \text{mK}$, and $T(10^{10}) \simeq 10 \text{mK}$, which is achievable in most dilution refrigerators.
\subsection{\label{subsec:level5} Qubit collision via $H_{xy}$ }
In the previous examples, we have been focusing on the excitation exchange type of Hamiltonian $H_{ee}$ (Eq.\,(\ref{eq:eq4})). It has a nice property that it keeps the total number of excitations conserved. This makes the structures of Q-state and $\square$-state invariant when they interact with a new ground state qubit. However, this property no longer holds for $H_{xy}$, which contains a $\sigma_{B,+} \otimes \sigma_{C,+}+\sigma_{B,-} \otimes \sigma_{C,-}$ term. This means that even if one starts with the $Q$-state, upon its interaction with another qubit, there will generally comes out a nonzero $\rho_{11}$ and $\rho_{14}$ term, making the concurrence contain a $-\sqrt{\rho_{11}\rho_{44}}$ or $-\sqrt{\rho_{22}\rho_{33}}$ term. As has been analyzed in the previous subsection, such a negative term always makes it harder to form entanglement nonlocally. Actually, the pairwise entanglement formed under this Hamiltonian tends to be highly localized, meaning that either entanglement cannot form nonlocally or it dies off quickly. In Fig.\,\ref{fig:epsart7}, we show a few examples of the pairwise entanglement under this Hamiltonian, which belongs to the 1st model in Table \ref{tab:table1}.

\subsection{\label{subsec:level6} Qubit cooled down: Many-to-one collision}
\begin{figure}[!tbp]
\centering
\begin{tikzpicture}[x=0.75pt,y=0.75pt,yscale=-1,xscale=1]
\node (Al) at (50,0) {$A_l$};
\node (B) at (140,0) {$B$};
\node (Ci) at (200,-40) {$C_i$};
\node (Cj) at (230,0) {$C_j$};
\node (Ck) at (200,40) {$C_k$};

\draw[->] (Ci) edge (B);
\draw[->] (Cj) edge (B);
\draw[->] (Ck) edge (B);
\draw[dashed]  (Ci) --(Al);
\draw[dashed]  (Ck) --(Al);
\draw[dashed]  (Ci) to (Cj);
\draw[dashed]  (Cj) to (Ck);
\draw[dashed]  (Ci) to (Ck);
\draw[dashed]  (Al) to[out=-50,in=-130]  (Cj);
\draw [decorate,decoration={coil,aspect=0,segment length=4pt,amplitude=0.6pt}]   (Al) -- (B);

\end{tikzpicture}
\caption{\label{fig:epsart9}General entanglement dynamics of the many-to-one collision model. A group of disentangled and non-interacting qubits, labeled as C$_1$, C$_2$, \dots, C$_n$ (only three drawn), simultaneously interact with qubit B, which was already entangled with A$_1$, A$_2$, \dots, A$_n$ (only one drawn) before collision. There is no A-B interaction. Note that even though there are no interaction between C$_i$ and C$_j$, they may become entangled post-collision.}
\end{figure}
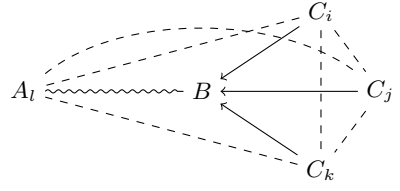

In the above, we have showed the entanglement dynamics where each collision involves only two qubits, i.e. one-to-one collision. In fact, the model can be extended to allow for many-to-one collision, where a group of disentangled and non-interacting qubits simultaneously interact with an old qubit; see Fig.\,\ref{fig:epsart9}. Note that even though the group of qubits are not interacting with each other, they may nevertheless become entangled post-collision because of their simultaneous interaction with the same qubit. As one may expect, the dynamics of many-to-one collision will be more complicated than that of one-to-one collision, and analytical results will be more limited. Here we show a special case where we can solve for the tangle $\tau$ analytically after each collision.

Suppose all qubits have the same frequency $\omega$, and that the interaction Hamiltonian is of the type $H_{ee}$ for each B--C$_i$ collision. The total Hamiltonian is given by
\begin{equation} \label{eq:eq15}
\begin{aligned}
	H_\text{tot}&=\sum_i\Omega_i\left(\sigma_{B,+} \otimes \sigma_{C_i,-}+h.c.\right)+\sum_{i=\text{all qubits}}\frac{1}{2}{\omega}\sigma_{i,z},
	\end{aligned}
\end{equation}
where $\Omega_i$ is the interaction strength between B--C$_i$ interaction. Let $\rho_{AB}$ be initially at Q-state, and all new qubits C$_i$ are prepared at $\ket{0}$. Then it can be shown that the post-collision entanglement as measured by tangle $\tau$ for each qubit pair can be given by
\begin{equation} \label{eq:eq16}
\begin{aligned}
	&\tau^\prime_{BC_i}=\frac{\Omega_i^2}{|\vec \Omega|^2} \rho_{33}^2 \sin^2(2t|\vec \Omega|),\;
	\tau^\prime_{C_i C_j}=\frac{\Omega_i^2 \Omega_j^2}{|\vec \Omega|^4} \rho_{33}^2 \sin^4(t|\vec \Omega|),\\
		&\tau^\prime_{A_lC_i}=\frac{\Omega_i^2}{|\vec \Omega|^2} \tau_{A_lB} \sin^2(t|\vec \Omega|),\;
			\tau^\prime_{A_lB}=\tau_{A_lB} \cos^2(t|\vec{\Omega}|),
\end{aligned}
\end{equation}
where $|\vec{\Omega}|=\sqrt{\sum_i \Omega_i^2}$, and $\rho_{33}=\bra{1}\rho_B\ket{1}$, the same as previously defined. The results show that there exists a competition in generating new entanglement ---  interactions with a larger interaction strength lead to stronger new entanglements. Note again how the nonlocal change of entanglement about qubit A$_l$ is a conserved quantity in this special example, i.e. 
$
	\tau^\prime_{A_l B}+\sum_i \tau^\prime_{A_l C_i} = \tau_{A_l B}.
$
It is also interesting to note that the newly formed entanglement between C$_i$ and C$_j$ is much weaker than $\tau^\prime_{BC_i}$ and $\tau^\prime_{A_l C_i}$.
In the Appendix \ref{sec:appendix4}, we show detailed derivations of the above results, including a more general case where $\rho_{AB}$ is initially at the X-state.

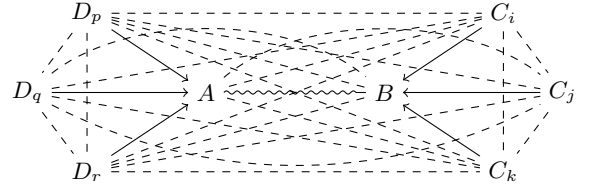
\begin{figure}[!t]
\centering
\begin{tikzpicture}[x=0.75pt,y=0.75pt,yscale=-1,xscale=1]
\node (Al) at (50,0) {$A$};
\node (B) at (140,0) {$B$};
\node (Ci) at (200,-40) {$C_i$};
\node (Cj) at (230,0) {$C_j$};
\node (Ck) at (200,40) {$C_k$};
\node (Di) at (-10,-40) {$D_p$};
\node (Dj) at (-40,0) {$D_q$};
\node (Dk) at (-10,40) {$D_r$};

\draw[->] (Ci) edge (B);
\draw[->] (Cj) edge (B);
\draw[->] (Ck) edge (B);
\draw[->] (Di) edge (Al);
\draw[->] (Dj) edge (Al);
\draw[->] (Dk) edge (Al);
\draw[dashed]  (Ci) --(Al);
\draw[dashed]  (Ck) --(Al);
\draw[dashed]  (Di) --(B);
\draw[dashed]  (Dk) --(B);
\draw[dashed]  (Ci) to (Cj);
\draw[dashed]  (Cj) to (Ck);
\draw[dashed]  (Ci) to (Ck);
\draw[dashed]  (Di) to (Dj);
\draw[dashed]  (Dj) to (Dk);
\draw[dashed]  (Di) to (Dk);

\draw[dashed]  (Di) to (Ck);
\draw[dashed]  (Di) to (Cj);
\draw[dashed]  (Di) to (Ci);
\draw[dashed]  (Dk) to (Ci);
\draw[dashed]  (Dk) to (Cj);
\draw[dashed]  (Dk) to (Ck);
\draw[dashed]  (Dj) to (Ci);
\draw[dashed]  (Dj) to (Ck);

\draw[dashed]  (Al) to[out=-50,in=-130]  (Cj);
\draw[dashed]  (Dj) to[out=-50,in=-130]  (B);
\draw[dashed]  (Dj) to[out=35,in=145]  (Cj);
\draw [decorate,decoration={coil,aspect=0,segment length=4pt,amplitude=0.6pt}]   (Al) -- (B);

\end{tikzpicture}
\caption{\label{fig:epsart10}Two groups of disentangled and non-interacting qubits, labeled as C$_1$, C$_2$, \dots, C$_n$, and D$_1$, D$_2$, \dots, D$_n$ (only three from each group are shown), simultaneously interact with qubit A and B, respectively. Qubit B and A are entangled pre-collision but do not interact with each other.}
\end{figure}

On the other hand, one can also consider the model where two groups of non-interacting and disentangled qubits simultaneously interacting separately with two entangled qubits, A and B (Fig.\,\ref{fig:epsart10}). This model can be understood as a process where a spatially separated entangled qubit pair experiences decoherence because of their interactions with their local environment. The system will also become an entanglement quilt after collisions. To analytically solve for this model, note that the Hamiltonian $H_{DA}$ of D--A collisions commutes with the Hamiltonian $H_{BC}$ of B--C collisions. Hence the total time evolution operator $U_\text{tot}$ could be decomposed into $U_\text{tot}=U_{DA}U_{BC}$, meaning that this model can be reduced to the model described in Eq.\,(\ref{eq:eq15}), and one only needs to apply the Eq.\,(\ref{eq:eq16}) twice, respectively, to each subsystem.

\section{\label{sec:level4}Discussions}
\subsection{\label{subsec:level06}New qubits beyond $\ket{0}$ and $\ket{1}$}
In the previous sections, we have only considered the case where the new qubits are either prepared as $\ket{0}$ or $\ket{1}$. One may wonder if it is possible to generalize and extend the model to include superimposed new qubits, such as $\frac{1}{\sqrt{2}}(\ket{0}+\ket{1})$. However, it turns out that the concurrence of interacting qubits in more general setups exhibits a complicated expression, even for the simplest case. To understand, consider two qubits A and B, which are initially prepared at
\begin{equation} \label{eq:eq17}
\begin{aligned}
	\ket{\psi_A}&=e^{i \phi_1} \cos \theta_{1}|1\rangle+\sin \theta_{1}|0\rangle\\
	\ket{\psi_B}&=e^{i \phi_2} \cos \theta_{2}|1\rangle+\sin \theta_{2}|0\rangle.
\end{aligned}
\end{equation}
Assuming they have the same frequency $\omega$, and they interact with each other via the excitation-exchange type of Hamiltonian $H_{ee}$ for time $t$. Their tangle after the collision can be given by
\begin{widetext}
\begin{equation} \label{eq:eq18}
	\tau^\prime_{AB}=\sin ^{2}(t)[\underbrace{b \,c \sin (\phi) \sin  (2 t_{} )-c^{2} \cos ^{2} (t_{} ) \sin ^{2} \phi}_{=0 \text { if } \phi=0}+(1-a)^{2}-b^{2} \sin ^{2} t_{}],
\end{equation}
\end{widetext}
where we have taken $\Omega=1$ and defined
\begin{equation} \label{eq:eq19}
	\begin{aligned}
		a&=\cos  (2 \theta_{1} ) \cos  (2 \theta_{2} )\\
		b&=\cos  (2 \theta_{1} )-\cos  (2 \theta_{2} )\\
		c&=\sin  (2 \theta_{1} ) \sin  (2 \theta_{2} )\\
		\phi&=\phi_{1}-\phi_{2}.
	\end{aligned}
\end{equation}
To verify this result, recall that one needs to calculate the eigenvalues $\lambda_i$ of $\rho^\prime_{AB} \tilde{\rho}^\prime_{AB}$ to obtain $C^\prime_{AB}$ [Eq.\,(\ref{eq:eq1})]. By proving $\operatorname{Tr}{\rho^\prime_{AB} \tilde{\rho}^\prime_{AB}}=\lambda_1+\lambda_2+\lambda_3+\lambda_4=\tau^\prime_{AB}$ and showing $\tau^\prime_{AB}$ is one of the eigenvalues of $\rho^\prime_{AB} \tilde{\rho}^\prime_{AB}$, one concludes $\lambda_{2}=\lambda_{3}=\lambda_{4}=0$ (since $\lambda_i \geq0$) and therefore $\lambda_1=\tau^\prime_{AB}$.

Lastly, we note that the above expression reduces to $\tau^\prime_{AB}=\sin^2(2t)$ in the case of $\ket{\psi_A}=\ket{1}$ and $\ket{\psi_B}=\ket{0}$. Because of the complexity of Eq.\,(\ref{eq:eq18}), further generalizations to multiple collisions are nontrivial.
\subsection{\label{subsec:level7}Generalization to interactions between old qubits}
Another direction of generalization is to consider the interactions between old qubits and study how such interactions affect the pairwise entanglement of each qubit pairs. However, this can be quite nontrivial. Since in the general case, each of the two interacting qubits is already entangled with many other qubits before collisions, a correct treatment would require one to take all entangled qubits into consideration, meaning that the dimension of the involved Hilbert space will scales exponentially with the number of qubits. For a system which is initially prepared as an entanglement quilt, this essentially requires one to solve for the Sch\"{o}rdinger equation for the whole system.

There is an imperfect workaround that works only for a limited number of collisions. Consider, for example, an entanglement quilt, where B collides with C, and they are both entangled to A$_i$ before the collision. One can calculate $\rho^\prime_{A_iBC}$ for each $i$, which only involves an 8-dimensional Hilbert space each time. After obtaining $\rho^\prime_{A_iBC}$, one can solve for $\rho^\prime_{A_i B}$, $\rho^\prime_{A_i C}$ and $\rho^\prime_{B C}$ to calculate their concurrence. Note that the entanglement between A$_i$ and A$_j$ does not change.

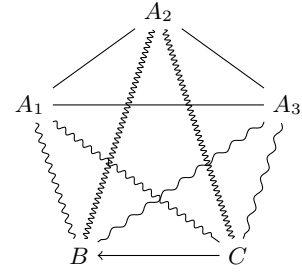
\begin{figure}[!tp]
\begin{center}
\begin{tikzpicture}[x=0.75pt,y=0.75pt,yscale=-1,xscale=1]
\node (A1) at (-25,-76) {$A_1$};
\node (A2) at (40,-123) {$A_2$};
\node (A3) at (105,-76) {$A_3$};
\node (B) at (0,0) {$B$};
\node (C) at (80,0) {$C$};
\draw[->] (C) edge (B);
\draw    (A1) -- (A2);
\draw    (A2) -- (A3);
\draw    (A1) -- (A3);

\draw    [decorate,decoration={coil,aspect=0,segment length=4pt,amplitude=1pt}](A1) -- (B);
\draw    [decorate,decoration={coil,aspect=0,segment length=2pt,amplitude=1pt}](A2) -- (B);
\draw    [decorate,decoration={coil,aspect=0,segment length=8pt,amplitude=1pt}](A3) -- (B);

\draw    [decorate,decoration={coil,aspect=0,segment length=4pt,amplitude=1pt}](A1) -- (C);
\draw    [decorate,decoration={coil,aspect=0,segment length=2pt,amplitude=1pt}](A2) -- (C);
\draw    [decorate,decoration={coil,aspect=0,segment length=8pt,amplitude=1pt}](A3) -- (C);
\end{tikzpicture}
\end{center}
\caption{\label{fig:epsart11} Two old qubits B and C, which are previously entangled with other qubits labeled as A$_1$, A$_2$, \dots, A$_n$ (only three displayed), collides with each other via excitation exchange Hamiltonian $H_{ee}$. The wavy lines denote the entanglement between A$_i$ and B (or C) has changed because of the B--C interaction. Here we have used three different wavy lines, and for each kind, the total entanglement change is zero. For instance, the entanglement change between A$_1$ and B is exactly compensated by the entanglement change between A$_1$ and C, such that if one increases then the other decreases by the same amount, and vice versa. The entanglement between B--C may increase or decrease, depending on both the collision time $t$ and pre-collision state of $\rho_{BC}$. The entanglement between A$_i$ and A$_j$ is unchanged, hence labeled by straight lines.}
\end{figure}

However, in the next time step, suppose A$_1$ collides with A$_2$. There is no difficulties in calculating $\rho^{\prime\prime}_{A_1 A_2 A_i}$ for all $i \geq 3$. Yet to know the entanglement change between e.g. A$_1$--B and A$_1$--C, one needs to calculate $\rho^{\prime\prime}_{A_1 A_2 B}$ and $\rho^{\prime\prime}_{A_1 A_2 C}$, which in turn requires one to know the pre-collision density matrices $\rho^{\prime}_{A_1 A_2 B}$ and $\rho^\prime_{A_1 A_2 C}$. To correctly obtain $\rho^{\prime}_{A_1 A_2 B}$ and $\rho^\prime_{A_1 A_2 C}$, one needs to calculate $\rho^{\prime}_{A_1 A_2 B C}$ to perform a partial tracing, which involves a 16-dimensional Hilbert space. The similar process continues to occur in the subsequent collisions, and each time, in the worst case, requires the inclusion of one more qubit for calculation, doubling the dimension of the Hilbert space. After sufficiently many collisions, one essentially needs to include all qubits for calculations, which is equivalent to solving the Sch\"{o}rdinger equation for the whole system. Hence, the workaround allows one to obtain useful information only for a limited number of collisions.

Nevertheless, in the following, we show an example where the pairwise entanglement dynamics can be obtained despite the interactions between correlated qubits. Consider the example in Section \ref{subsec:level3}. As has been mention before, it can be shown that the state of the total system can always be given by
\begin{equation} \label{eq:eq20}
	\ket{\psi}=a_1 \ket{100\dots0}+ a_2\ket{010\dots0}+\dots +a_n \ket{000\dots1},
\end{equation}
where $a_i \in \mathbb{C}$ are local factors satisfying the normalization condition $\sum_{i=1}^n |a_i|^2=1$. The reduced density matrix between the $i$th and $j$th qubit can be given by
\begin{equation} \label{eq:eq21}
	\rho_{ij}=\begin{pmatrix}
	0&0&0&0\\
	0&|a_i|^2&a_i a^*_j&0\\
	0&a^*_i a_j&|a_j|^2&0\\
	0&0&0&1-|a_i|^2-|a_j|^2
\end{pmatrix},
\end{equation}
which is a Q-state, meaning that the tangle between the $i$th and $j$th qubit is given by $\tau_{ij}=4|a_i|^2 |a_j|^2 $. Now consider two arbitrary old qubits, say the $k_1$th and $k_2$th qubit, which are already entangled with other qubits, collides with each other via the excitation exchange Hamiltonian $H_{ee}$. The post-collision state for the total system can be given by
\begin{equation} \label{eq:eq22}
	\ket{\psi^\prime}=a^\prime_1 \ket{100\dots0}+ a^\prime_2\ket{010\dots0}+\dots +a^\prime_n \ket{000\dots1},
\end{equation}
with $a^\prime_1=a_1$, $a^\prime_2=a_2$, \dots, except that 
\begin{equation} \label{eq:eq23}
\begin{aligned}
	a^\prime_{k_1}&=a_{k_1} \cos(\Omega t)-i a_{k_2} \sin(\Omega t)\\
	a^\prime_{k_2}&=a_{k_2} \cos(\Omega t)-i a_{k_1} \sin(\Omega t).
\end{aligned}
\end{equation}
Note that $|a^\prime_{k_1}|^2+|a^\prime_{k_2}|^2=|a_{k_1}|^2+|a_{k_2}|^2$, which is a conserved quantity. This means that if $|a^\prime_{k_1}|^2$ increases with respect to $|a_{k_1}|^2$, then $|a^\prime_{k_2}|^2$ will decrease with respect to $|a_{k_2}|^2$, and vice versa. By making use of Eqs.\,(\ref{eq:eq21})-(\ref{eq:eq23}), one can solve for the post-collision reduced density matrices of $\rho^\prime_{k_1 k_2}, \rho^\prime_{k_1 i}, \rho^\prime_{k_2 i}$, and $\rho^\prime_{ij}$, where $i,j \notin \{k_1, k_2\}$. As all of these post-collision density matrices remain to be Q-states, the post-collision tangle can be given by,
\begin{equation} \label{eq:eq24}
\begin{aligned}
	\tau^\prime_{ij}&=\tau_{ij}\\
	\tau^\prime_{k_1 i}&=4 |a^\prime_{k_1}|^2 |a_i|^2\\
	\tau^\prime_{k_2 i}&=4 |a^\prime_{k_2}|^2 |a_i|^2\\
	\tau^\prime_{k_1 k_2}&=4 |a^\prime_{k_1}|^2 |a^\prime_{k_2}|^2.
\end{aligned}
\end{equation}
The results suggest that because of the interaction between the old qubits $k_1$ and $k_2$, the entanglement between the $k_1$th ($k_2$th) qubit and $i$th qubit has changed in such a way that if their entanglement increases by certain amount as measured by tangle, then the entanglement between the $k_2$th ($k_1$th) and $i$th qubit decreases by exactly the same amount, and vice versa (Fig.\,\ref{fig:epsart11}). On the other hand, the entanglement change between the $k_1$th and $k_2$th qubit has no definite direction --- it may increase or decrease, depending on the interaction time and pre-collision value of $a_{k_1}$ and $a_{k_2}$.

\begin{figure}[!tp]
\includegraphics[width=0.48\textwidth]{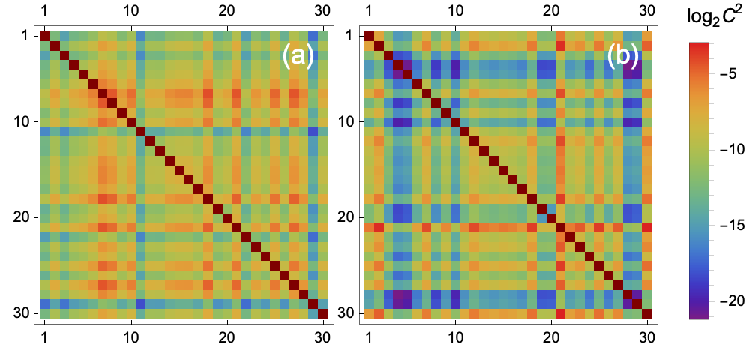}
\caption{\label{fig:epsart12} The pairwise entanglement of 30 qubits after $10^7$ collisions via excitation exchange type of Hamiltonian $H_{ee}$, with collisions between entangled qubits allowed. Panel (a): each time, two random qubits collides with each other; panel (b): each time, a random qubit collides with the first qubit. Repeated simulations show no noticeable visual differences regarding the pairwise entanglement between the two collisions schemes. We take the $\Omega=1$ and collision time $t$ to be a random value between 0 and $2\pi$ for each collision. Initial state: the first qubit at excited state and all other qubits at ground state.
}
\end{figure}

In Fig.\,\ref{fig:epsart12}, we show the heat map of pairwise entanglement of 30 qubits after $10^7$ random collisions, with collisions between entangled qubits allowed and collision time $t$ being a random number between $0$ and $2\pi$ (we set $\Omega=1$). The results can be regarded as an excited qubit being cooled down by a small cold bath and the whole system thermalizes after a sufficient number of qubit-qubit interactions.

\subsection{Implications for condensed matter systems}{\label{subsec:level10}
For any condensed matter systems $H=\sum_k H_k$, its evolution can always be broken into small time pieces $\Delta t =t/N$ by the Trotterization formula,
\begin{equation}
	e^{-i H t}=e^{-i\Sigma_{k} H_{k} t} \approx\left(\prod_{k} e^{-i H_{k} \frac{t}{N}}\right)^{N},
\end{equation}
which is actually how condensed matter systems can be simulated in quantum computers \cite{smith2019simulating}. The Trotterization formula bridges the many-body and few-body physics by allowing one to model the evolution of a many-body system as a series of fast quantum collision processes. That is, the many-body system evolves as if during each short time interval $\Delta t$, two particles collide (interact) with each other and all other particles do not interact with any particles. This picture coincides with quantum collision models, suggesting that we may apply some of the results obtained in this work to certain condensed matter system. 

For instance, in Ref.\,\cite{PhysRevA.64.012313}, it is shown that in the quantum Heisenberg XY model, if the total excitation is one (so that the wave function is a W-like state), then the entanglement between an arbitrary pair of qubits will generally be nonzero. However, we have shown that the pairwise entanglement in W-like state is sensitive to excitations because of the transition from Q-state to $\phi$-state. This suggests that the entanglement behavior of this model with higher excitations will be quite different from the case with a single excitation.

On the other hand, for condensed matter systems where the total number of excitations is not conserved and local interactions can be described by $H_{xy}$, the reduced density matrix of an arbitrary qubit pair will in general be at least an X-state, if its structure is not more complex. This suggests that for such systems, long-range entanglement between remote qubit pairs can usually not survive, and only local entanglement, such as the entanglement between nearest-neighbor qubit pairs, can exist. (See, for example, Refs.\,\cite{PhysRevA.66.032110} and \cite{PhysRevA.69.022304}).

\subsection{Entanglement quilts do not have to be W-like states}{\label{subsec:level9}
Although all W-like states are entanglement quilts, the converse is not true: Entanglement quilts do not have to be W-like states. To demonstrate, consider a three-qubit system $\ket{\psi}_{ABC}=\ket{100}$. Let B collides with A, followed by C colliding with B, for time $t$, respectively, via the $H_{xy}$ type of Hamiltonian. The total Hamiltonian during the collision is given by
\begin{equation}
	H_\text{AB,tot}=\sqrt{\Omega} \sigma_{A,x} \otimes \sigma_{B,x} + \sum_{i=A,B,C} \frac{1}{2}\omega \sigma_{i,z},
\end{equation}
and similarly for $H_\text{BC,tot}$. If we take $\Omega=\omega=1$, the post-collision state can be given by
\begin{equation}
\begin{aligned}
	\ket{\psi^\prime}_{ABC}&=-\frac{ie^{-it}}{\sqrt{2}} \sin(t) \sin{(\sqrt{2}t})\ket{111}+i \cos^2(t)\ket{100}\\&+\frac{1}{2}e^{-it}\sin (t) (2 \cos{(\sqrt{2}t)+i \sqrt{2}\sin{(\sqrt{2}t)}})\ket{010}\\&+\cos (t) \sin (t) \ket{001}.
	\end{aligned}
\end{equation}
This state is not a W-like state but can be shown to be an entanglement quilt if one takes, e.g. $t=\pi/3$.

Another family of states which are entanglement quilts are Dicke states. Consider $N$-particle Dicke states with $n$ spin-ups,
\begin{equation}
	|D_{N,n}\rangle=\binom{N}{n}^{-1 / 2} \sum_{\mathcal{P}} \mathcal{P}_{}\left(|0\rangle^{\otimes(N-n)} \otimes|1\rangle^{\otimes n}\right),
\end{equation}
where $\mathcal{P}$ denotes all permutations of $\ket{0}$'s and $\ket{1}$'s. The reduced density matrices are the same for all qubit pairs, which is a $\phi$-state, and is given by \cite{Xi_2012}
\begin{equation}
\begin{aligned}
	\rho_{11}&=\frac{n(n-1)}{N(N-1)}\\
	\rho_{12}&=\rho_{21}=\rho_{22}=\rho_{33}=\frac{n(N-n)}{N(N-1)}\\
	\rho_{44}&=\frac{(N-n)(N-n-1)}{N(N-1)}.
\end{aligned}
\end{equation}
It can be shown that the concurrence $C$ is always a positive number, equal to $2 (\rho_{12}-\sqrt{\rho_{11}\rho_{44}})$.

\subsection{Entanglement between arbitrary subsystems of an entanglement quilt}{\label{subsec:level8}
Although the full properties of entanglement quilts are still to be explored, here we speculate about a possible property: For a many-body system that is an entanglement quilt, every subsystem of it is entangled to every other subsystem (the two subsystems of interest must not share a particle, of course). For instance, this means that in Fig.\,\ref{fig:epsart11}, qubit B and C as a whole (denoted as \{B,C\}) is entangled with arbitrarily picked $m$ qubits A as a whole, $\{\mathrm{A}_{i_1}, \mathrm{A}_{i_2}, \dots, \mathrm{A}_{i_m}\}$. If this speculation is true, then such property suggests another possible form of maximal entanglement, where the entanglement is \textit{maximally distributed} among subsystems, distinct from that of Greenberger–Horne–Zeilinger state (GHZ state).

The motivation of the speculation comes from the \textit{monogamy of entanglement}, which states that for a many-qubit system A$_1$, A$_2$, $\dots$, A$_n$, its tangle follows the Coffman–Kundu–Wootters (CKW) inequality \cite{PhysRevA.61.052306, PhysRevLett.96.220503},
\begin{equation} \label{eq:eq30}
	\sum_{k=2}^{n} \tau\left(\rho_{A_{1} A_{k}}\right) \leq \tau\left(\rho_{A_{1}\left\{A_{2} \ldots A_{n}\right\}}\right).
\end{equation}
Hence, for an entanglement quilt this implies each qubit (say, A$_1$) is entangled to each of the subsystems, $\{\mathrm{A}_{i_1}, \mathrm{A}_{i_2}, \dots, \mathrm{A}_{i_m}\}$ where $i_j \neq 1$ and $m < n$ is arbitrarily determined. If a similar inequality can be proved for qudits, then the speculation will automatically become true as multiple qubits as a whole can be regarded as a qudit. 

To see this, consider an entanglement quilt wherein a total number of $(n+m)$ qubits are labeled as $\mathrm{A}_1, \mathrm{A}_2, \dots, \mathrm{A}_n, \mathrm{B}_1, \mathrm{B}_2, \dots, \mathrm{B}_m$. Note that the label is arbitrary and one can always change it if needed. From the above discussions, we know that $\{\mathrm{B}_1, \mathrm{B}_2, \dots, \mathrm{B}_m\}$ is entangled with A$_i$ for all $i \in \mathbb{Z}_n$. We can regard $\{\mathrm{B}_1, \mathrm{B}_2, \dots, \mathrm{B}_m\}$ as a $2^m$-dimensional qudit and relabel it as $\mathrm{\tilde{B}}$. If the generalized CKW inequality holds,
\begin{equation} 
	\sum_{k=1}^{n^\prime} \tau\left(\rho_{\tilde{B} A_{i_k}}\right) \leq \tau\left(\rho_{\tilde{B}\left\{A_{i_1}A_{i_2} \ldots A_{i_{n^\prime}}\right\}}\right),
\end{equation}
where $n^\prime \leq n$, then every subsystem $\{\mathrm{B}_1, \mathrm{B}_2, \dots, \mathrm{B}_m\}$ are entangled with every other subsystem $\{A_{i_1}A_{i_2} \ldots A_{i_{n^\prime}}\}$.

\section{\label{sec:level5}Conclusion}
In this article, we propose a framework for studying the pairwise entanglement dynamics of a family of quantum collision models. By utilizing diagrammatic descriptions, we are able to monitor and understand intuitively the entanglement dynamics of many-body systems on a collision-by-collision and frame-by-frame basis. This approach illustrates clearly the cause-and-effect relationships of entanglement generation, propagation, and decay, which have been largely obscured in previous studies due to the many degrees of freedom involved. In the multiple examples we show, we identify a type of genuine multipartite entanglement called entanglement quilt, where every qubit is entangled with every other qubit. We find that entanglement quilt can be hypersensitive to local excitation fluctuations, such that even a single excited qubits can destroy the entanglement quilt. This is because the alteration of the structures of the reduced two-qubit density matrices, which makes it generally harder to be an entangled state. This finding brings insight into the lack of long-range entanglement in typical condensed matter systems. Moreover, previous studies on entanglement and entanglement dynamics of spin chains have mostly focused on special cases, such as ground states, thermal states, the ground state with some excitation(s) added \cite{PhysRevA.69.034304} (e.g. $\ket{100\dots00}$), and ground states with an initial Bell state $\ket{\Phi}$ added (e.g. $\ket{\Phi}\otimes\ket{0\dots0}$) \cite{PhysRevA.69.034304, PhysRevA.69.022304}. Our findings suggest that the entanglement and its dynamics could be very different beyond these special cases, meriting further investigation.
\section{Acknowledgement}
This work was supported by the U.S. Army Research Office under grant W911NF-22-1-0258. We also thank 
Chapman University for research support.
 \appendix
 \begin{widetext}
 \bigskip

\setcounter{equation}{0}
\setcounter{figure}{0}
\setcounter{table}{0}
\makeatletter
\renewcommand{\theequation}{A\arabic{equation}}
\renewcommand{\thefigure}{A\arabic{figure}}
\renewcommand{\thetable}{A\arabic{table}}

\section{Recurrence relations of the 1st models in Table \label{sec:appendix1}}
Below in Table \ref{tab:appendixtable3} we provide the recurrence relations for the density matrix entries in the 1st model in Table \ref{tab:table1}.
\begin{table}[!h]
  \caption{Recurrence relations of the reduced density matrices $\rho_{AB}$, $\rho_{BC}$, $\rho_{AC}$ of the 1st model in Table \ref{tab:table1}, where the new qubit $C$ are initially at $\ket{0}$ or $\ket{1}$ and $\rho_{AB}$ has the X structure. Note that here, $\rho_{ij}=[\rho_{AB}]_{ij}$ whereas $\rho^\prime_{ij}=[\rho^\prime_{XY}]_{ij}$ where $XY=AB, AC$, or $BC$.}
  \renewcommand{\arraystretch}{1.45}
    \centering
    \begin{tabular}{ccccc}
    \toprule
        &&&\\[-2em]
    $\rho_C$&$[\rho^\prime_{XY}]_{ij}$&$\rho^\prime_{AB}$&$\rho^\prime_{BC}$ & $\rho^\prime_{AC}$     \\
    \midrule
    &&&\\[-1.9em]
    \multirow{6}{*}{$\ket{0}\bra{0}$}&$\rho^\prime_{11}$&$\rho_{11}\beta^-_{\text{c}}+\rho_{22}\beta^+_s$     &  $(\rho_{22}+\rho_{44})\beta^+_{s}$  &$\rho_{11}\beta^-_s+\rho_{22}\beta^+_s$   \\ 
  &$ \rho^\prime_{22}$&$\rho_{22}\beta^+_{c}+\rho_{11}\beta^-_s$     &   $(\rho_{11}+\rho_{33})\beta^-_{c}$    &   $\rho_{11}\beta^-_{c}+\rho_{22}\beta^+_c$   \\
   &$ \rho^\prime_{33}$&$\rho_{33}\beta^-_c+\rho_{44}\beta^+_s$     & $(\rho_{11}+\rho_{33})\beta^-_{s}$     &    $\rho_{33}\beta^-_{s}+\rho_{44}\beta^+_{s}$    \\
    &$\rho^\prime_{44}$&$\rho_{44}\beta^+_c+\rho_{33}\beta^-_s$     &  $(\rho_{22}+\rho_{44})\beta^+_{c}$    &   $\rho_{33}\beta^-_{c}+\rho_{44}\beta^+_c$  \\
    &$\rho^\prime_{23}$&$e^{-i\omega_A t}(e^{2 i \theta} \rho_{14} \sqrt{\beta_{s}^{-} \beta_{s}^{+}}+\rho_{23} a b^{*})$     &       $[\rho^\prime_{BC}]^0_{23}$&    $e^{-i \omega_A t}(\rho_{14} b e^{2 i \theta} \sqrt{\beta_{s}^{+}}+\rho_{23} a \sqrt{\beta_{s}^{-}})$    \\
   &$ \rho^\prime_{14}$&$e^{-i\omega_A t}(e^{-2 i \theta} \rho_{23} \sqrt{\beta_{s}^{-} \beta_{s}^{+}}+\rho_{14} a^* b)$     &       $[\rho^\prime_{BC}]^0_{14}$&    $e^{-i \omega_{A} t}(\rho_{14} a^{*} \sqrt{\beta_{s}^{-}}+\rho_{23} b^{*} e^{-2 i \theta} \sqrt{\beta_{s}^{+}})$  \\ 
   \midrule
      \multirow{6}{*}{$\ket{1}\bra{1}$}&$\rho^\prime_{11}$&$\rho_{11}\beta^+_{c}+\rho_{22}\beta^-_s$&$(\rho_{11}+\rho_{33})\beta_c^+$&$\rho_{11}\beta^+_c+\rho_{22}\beta^-_c$\\
   &$\rho^\prime_{22}$&$\rho_{22}\beta^-_c+\rho_{11}\beta^+_s$&$(\rho_{22}+\rho_{44})\beta^-_s$&$\rho_{11}\beta^+_s+\rho_{22}\beta^-_s$\\
   &$\rho^\prime_{33}$&$\rho_{33}\beta^+_c+\rho_{44}\beta^-_{s}$&$(\rho_{22}+\rho_{44})\beta^-_c$&$\rho_{33}\beta^+_c+\rho_{44}\beta^-_c$\\
   &$\rho^\prime_{44}$&$\rho_{44}\beta^-_{c}+\rho_{33}\beta^+_{s}$&$(\rho_{11}+\rho_{33})\beta^+_s$&$\rho_{33}\beta^+_s+\rho_{44}\beta^-_s$\\
   &$\rho^\prime_{23}$&$e^{-i\omega_A t}(e^{2 i \theta} \rho_{14} \sqrt{\beta_{s}^{-} \beta_{s}^{+}}+\rho_{23} a b^{*})$&$[\rho^\prime_{BC}]^1_{23}$&$e^{-i \omega_{A} t}(\rho_{14} b^{*} e^{2 i \theta} \sqrt{\beta_{s}^{+}}-\rho_{23} a \sqrt{\beta_{s}^{-}})$\\
   &$\rho^\prime_{14}$&$e^{-i\omega_A t}(e^{-2 i \theta} \rho_{23} \sqrt{\beta_{s}^{-} \beta_{s}^{+}}+\rho_{14} a^* b)$&$[\rho^\prime_{BC}]^1_{14}$&$e^{-i \omega_{A} t}(-\rho_{14} a^{*} \sqrt{\beta_{s}^{-}}+\rho_{23} b e^{-2 i \theta} \sqrt{\beta_{s}^{+}})$\\
   \bottomrule
    \end{tabular}%
  \label{tab:appendixtable3}%
\end{table}
\begin{equation}
\begin{aligned}
	[\rho^\prime_{BC}]^0_{23}&=-\Omega\left(\rho_{11}+\rho_{33}\right)\left(\frac{-i 2 \Delta_{-} \sin \left(2 t \Delta_{-}\right)+\left(\omega_{B}-\omega_{C}\right)\left(\cos \left(2 t \Delta_{-}\right)-1\right)}{\left(2 \Delta_{-}\right)^{2}}\right)\\
	[\rho^\prime_{BC}]^0_{14}&=e^{-2 i \theta} \Omega\left(\rho_{22}+\rho_{44}\right)\left(\frac{-i 2 \Delta_{+} \sin \left(2 t \Delta_{+}\right)+\left(\omega_{B}+\omega_{C}\right)\left(\cos \left(2 t \Delta_{+}\right)-1\right)}{\left(2 \Delta_{+}\right)^{2}}\right)\\
	[\rho^\prime_{BC}]^1_{23}&=\Omega\left(\rho_{22}+\rho_{44}\right)\left(\frac{-i 2 \Delta_{-} \sin \left(2 t \Delta_{-}\right)+\left(\omega_{B}-\omega_{C}\right)\left(\cos \left(2 t \Delta_{-}\right)-1\right)}{\left(2 \Delta_{-}\right)^{2}}\right)\\
	[\rho^\prime_{BC}]^1_{14}&=-e^{-2 i \theta} \Omega\left(\rho_{11}+\rho_{33}\right)\left(\frac{-i 2 \Delta_{+} \sin \left(2 t \Delta_{+}\right)+\left(\omega_{B}+\omega_{C}\right)\left(\cos \left(2 t \Delta_{+}\right)-1\right)}{\left(2 \Delta_{+}\right)^{2}}\right).
\end{aligned}
\end{equation}
In the above, we have defined
\begin{equation}
	\Delta_{\pm}=\sqrt{\Omega^{2}+\left[\frac{1}{2}\left(\omega_{B}\pm\omega_{C}\right)\right]^{2}}
\end{equation}
\begin{equation}
\begin{aligned}
	\beta_{c}^{\pm}&=\frac{\Omega^{2} \cos ^{2}\left(t \Delta_{\pm}\right)+\Delta_{\pm}^{2}-\Omega^{2}}{\Delta_{\pm}^{2}}\\
	\beta_{s}^{\pm}&=\frac{\Omega^{2} \sin ^{2}\left(t \Delta_{\pm}\right)}{\Delta_{\pm}^{2}},
	\end{aligned}
\end{equation}
which satisfies $\beta^\pm_c+\beta^\pm_s=1$, and 
\begin{equation}
\begin{aligned}
	a&=\left(i \cos \left(t \Delta_{+}\right)-\frac{\left(\omega_{B}+\omega_{C}\right) \sin \left(t \Delta_{+}\right)}{2 \Delta_{+}}\right)\\
	b&=\left(i \cos \left(t \Delta_{-}\right)+\frac{\left(\omega_{B}-\omega_{C}\right) \sin \left(t \Delta_{-}\right)}{2 \Delta_{-}}\right).
	\end{aligned}
\end{equation}
Note that if qubit C collides with qubit A instead of qubit B, one must transform $\rho_{AB}$ to $\rho_{BA}$ before calculations:
\begin{equation} \label{eq:eqa5}
	\rho_{AB}=\left(\begin{array}{cccc}\rho_{11} & 0 & 0 & \rho_{14} \\ 0 & \rho_{22} & \rho_{23} & 0 \\ 0 & \rho_{32} & \rho_{33} & 0 \\ \rho_{41} & 0 & 0 & \rho_{44}\end{array}\right)_{A B} \quad \Rightarrow \quad \rho_{BA}=\left(\begin{array}{cccc}\rho_{11} & 0 & 0 & \rho_{14} \\ 0 & \rho_{33} & \rho_{32} & 0 \\ 0 & \rho_{23} & \rho_{22} & 0 \\ \rho_{41} & 0 & 0 & \rho_{44}\end{array}\right)_{BA}
\end{equation}

 \setcounter{equation}{0}
\setcounter{figure}{0}
\setcounter{table}{0}
\makeatletter
\renewcommand{\theequation}{B\arabic{equation}}
\renewcommand{\thefigure}{B\arabic{figure}}
\renewcommand{\thetable}{B\arabic{table}}

\section{Recurrence relations of the 2-4th models in Table \label{sec:appendix2}}
Below in Table \ref{tab:appendixtable2} we provide the recurrence relations for the density matrix entries in the 2nd models in Table \ref{tab:table1}. As the 3rd and 4th model can be considered as a special case of the 2nd model, this table is also applicable to these models.

\begin{table}[!h]
  \caption{Recurrence relations of the reduced density matrices $\rho_{AB}$, $\rho_{BC}$, $\rho_{AC}$ of the 2nd (and hence, the 3rd and 4th) model in Table \ref{tab:table1}, where the new qubit $C$ are initially at $\ket{0}$ or $\ket{1}$ and $\rho_{AB}$ has the X structure. Note that here, $\rho_{ij}=[\rho_{AB}]_{ij}$ whereas $\rho^\prime_{ij}=[\rho^\prime_{XY}]_{ij}$ where $XY=AB, BC$, or $AC$.}
  \renewcommand{\arraystretch}{1.45}
    \centering
    \begin{tabular}{cccccccc}
    \toprule
    &&&&&&\\[-2em]
    &\multicolumn{3}{c}{$\rho_C=\ket{0}\bra{0}$}&\multicolumn{3}{c}{$\rho_C=\ket{1}\bra{1}$}\\  \cmidrule(lr){2-4} \cmidrule(lr){5-7}
        &&&&&&\\[-2em]
    &$\rho^\prime_{AB}$&$\rho^\prime_{BC}$ & $\rho^\prime_{AC}$  &$\rho^\prime_{AB}$&$\rho^\prime_{BC}$&$\rho^\prime_{AC}$   \\
    \midrule
    &&&&&&\\[-1.9em]
    $\rho^\prime_{11}$&$\rho_{11}\beta_{\text{c}}$     &  0     &   $\rho_{11}\beta_s$&$\rho_{11}+\rho_{22}\beta_s$&$\rho_{11}+\rho_{33}$& $\rho_{11}+\rho_{22}\beta_c$\\ 
  $ \rho^\prime_{22}$&$\rho_{22}+\rho_{11}\beta_{\text{s}}$     &   $(\rho_{11}+\rho_{33})\beta_c$    &   $\rho_{22}+\rho_{11}\beta_c$  &$\rho_{22}\beta_c$&$(\rho_{22}+\rho_{44})\beta_s$& $\rho_{22}\beta_s$ \\
   $ \rho^\prime_{33}$&$\rho_{33}\beta_{\text{c}}$     & $(\rho_{11}+\rho_{33})\beta_s$     &    $\rho_{33}\beta_s$  &$\rho_{33}+\rho_{44}\beta_s$&$(\rho_{22}+\rho_{44})\beta_c$& $\rho_{33}+\rho_{44}\beta_c$  \\
    $\rho^\prime_{44}$&$\rho_{44}+\rho_{33}\beta_{\text{s}}$     &  $\rho_{44}+\rho_{22}$    &   $\rho_{44}+\rho_{33}\beta_c$  &$\rho_{44}\beta_c$&0&$\rho_{44}\beta_s$\\
    $\rho^\prime_{23}$&$\rho_{23}\sqrt{\beta_{c}}e^{i(r+R)}$     &    $i(\rho_{11}+\rho_{33})\sqrt{\beta_s\beta_c}e^{-iR}$   &    $i \rho_{23}\sqrt{\beta_s}e^{ir}$   &$\rho_{23}\sqrt{\beta_{c}}e^{i(r+R)}$&$i(\rho_{22}+\rho_{44})\sqrt{\beta_c\beta_s} e^{-iR}$& $-i\rho_{23}\sqrt{\beta_s}e^{ir}$\\
   $ \rho^\prime_{14}$&$\rho_{14}\sqrt{\beta_{c}}e^{i(s-R)}$     &    0   &    $-i \rho_{14}\sqrt{\beta_s}e^{is}$  &$\rho_{14}\sqrt{\beta_{c}}e^{i(s-R)}$&0&$i\rho_{14}\sqrt{\beta_s}e^{is}$ \\ 
   \bottomrule
    \end{tabular}%
  \label{tab:appendixtable2}%
\end{table}
Note that $\rho^\prime_{41}=[\rho^\prime_{14}]^*$, $\rho^\prime_{32}=[\rho^\prime_{23}]^*$, and all other entries not listed above vanish. Here we have defined
\begin{equation}\label{eq:eqb1}
\begin{aligned}
	R&=\arctan \left(\frac{\delta}{\Delta} \tan \left(t \Delta\right)\right)\\
	r&=-\frac{1}{2}\left(\omega_{A}-\omega_{B}\right) t-\frac{1}{2}\left(\omega_{A}-\omega_{C}\right) t\\
	s&=-\frac{1}{2}\left(\omega_{A}+\omega_{B}\right) t-\frac{1}{2}\left(\omega_{A}+\omega_{C}\right) t\\
	\delta&=\frac{1}{2}(\omega_B-\omega_C),
\end{aligned}
\end{equation}
\begin{equation}\label{eq:eqb2}
\begin{aligned}
	\beta_s&=\frac{\Omega^{2} \sin ^{2} t \Delta}{\Delta^{2}}\\
	\beta_c&=1-\beta_s=\frac{\Omega^{2} \cos ^{2} t \Delta+\Delta^{2}-\Omega^{2}}{\Delta^{2}}\\
	\Delta&=\sqrt{\Omega^2+\delta^2},
	\end{aligned}
\end{equation}
If all qubits have the same frequency $\omega$, then $e^{iR}=1$ and $e^{iS}=e^{-i2\omega t}$. 

Note that if qubit C collides with qubit A instead of qubit B, one must apply Eq.\,(\ref{eq:eqa5}) before performing calculations. With Table \ref{tab:appendixtable2}, it is a straightforward task to calculate concurrence for each pairs.

 \setcounter{equation}{0}
\setcounter{figure}{0}
\setcounter{table}{0}
\makeatletter
\renewcommand{\theequation}{C\arabic{equation}}
\renewcommand{\thefigure}{C\arabic{figure}}
\renewcommand{\thetable}{C\arabic{table}}
\section{Recurrence relations of the 5th models in Table \label{sec:appendix3}}
\begin{table}[h]
\centering
 \caption{Recurrence relations of the reduced density matrices $\rho_{AB}$, $\rho_{BC}$, $\rho_{AC}$ of the 5th model in Table \ref{tab:table1}, where the new qubit $C$ are initially at $\ket{0}$ or $\ket{1}$ and $\rho_{AB}$ has the $\square$ structure. Note that here, $\rho_{ij}=[\rho_{AB}]_{ij}$ whereas $\rho^\prime_{ij}=[\rho^\prime_{XY}]_{ij}$ where $XY=AB, BC$, or $AC$.}
\begin{tabular}{cccc}
	\toprule
	$[\rho^\prime_{XY}]_{ij}$&$\rho^\prime_{AB}$&$\rho^\prime_{BC}$&$\rho^\prime_{AC}$\\
	\midrule
	$\rho^\prime_{22}$&$\rho_{22}$&$\rho_{33}\beta_c$&$\rho_{22}$\\
	$\rho^\prime_{33}$&$\rho_{33}\beta_c$&$\rho_{33}\beta_s$&$\rho_{33}\beta_s$\\
	$\rho^\prime_{23}$&$\rho_{23}\sqrt{\beta_c}e^{i(r+R)}$&$i\rho_{33}\sqrt{\beta_c \beta_s}e^{-iR}$&$i\rho_{23}\sqrt{\beta_s}e^{ir}$\\
	\bottomrule
	\label{tab:appendixtable4}
\end{tabular}
\end{table}
Note that $\rho^\prime_{32}=[\rho^\prime_{23}]^*$, $\beta_c, \beta_s$ are given by Eq.\,(\ref{eq:eqb2}), and $R$ and $r$ are given by Eq.\,(\ref{eq:eqb1}). The recurrence relations for $\rho^\prime_{24}$, $\rho^\prime_{42}$, $\rho^\prime_{34}$, $\rho^\prime_{43}$, and $\rho^\prime_{44}$ are actually irrelevant because they never play a role in calculating the concurrence. All other entries vanish.


 \setcounter{equation}{0}
\setcounter{figure}{0}
\setcounter{table}{0}
\makeatletter
\renewcommand{\theequation}{D\arabic{equation}}
\renewcommand{\thefigure}{D\arabic{figure}}
\renewcommand{\thetable}{D\arabic{table}}

\section{Derivations for many-to-one collision \label{sec:appendix4}}
Below we provide a detailed calculation for Eq.\,(\ref{eq:eq16}). Note that during a many-to-one collision, the Hamiltonian can be given by 
\begin{equation}
	H=\underbrace{\sum_{i} \Omega_{i}\left(\sigma_{B,+} \otimes \sigma_{C_{i},-}+\sigma_{B,-} \otimes \sigma_{C_{i},+}\right)}_{H_1}+\underbrace{\frac{1}{2}\omega \sum_{i=\text{all qubits}}\sigma_{i,z}}_{H_{2}},
\end{equation}
where multiple new qubits, labeled as C$_i$, collides simultaneously with the same old qubit, labeled as B. The old qubit B is entangled with other qubits, labeled as A$_i$, before B-C$_i$ collision (Fig.\,\ref{fig:epsart9}). We have assumed that all qubits have the same qubit frequency $\omega$, and the interaction strength between B-C$_i$ is $\Omega_i$. Since $[H_1, H_2]=0$, we have $U_\text{tot}=U_1U_2$, where $U_i=\exp(-i H_i t)$ with $i=1,2$. This means that we can go to the rotating frame to cancel out $U_2$ and keep only the $H_1$ term. After all, $H_2$ is a local Hamiltonian so that it contributes nothing to the entanglement changes.

To proceed, we need to calculate post-collision density matrix $\rho^\prime_{ABC_1C_2 \hdots C_n}$ and then perform partial tracing to obtain $\rho_{AB},\rho_{BC_i}, \rho_{AC_i}$ and $\rho_{C_i C_j}$. This is sufficed by calculating $U_1=\exp(-i H_1 t)$. To calculate $U_1$, we make use of Taylor expansion $U_1=\sum_{i=1}^\infty \frac{1}{i!}(-i H_i t)^i$. Note that
\begin{equation}
	\begin{aligned}
H_1^2&=\sum_{i,j}\Omega_i\Omega_j(\sigma_{B,+}\otimes \sigma_{C_i,-}+\sigma_{B,-}\otimes \sigma_{C_i,+})(\sigma_{B,+}\otimes \sigma_{C_j,-}+\sigma_{B,-}\otimes \sigma_{C_j,+})\\
&=\sum_{i,j} \Omega_{i}\Omega_{j}(\underbrace{\sigma^2_{B,+}\otimes \sigma_{C_i,-} \sigma_{C_j,-}}_{0}+\sigma_{B,+}\sigma_{B,-}\otimes \sigma_{C_i,-}\sigma_{C_j,+}+\sigma_{B,-}\sigma_{B,+}\otimes \sigma_{C_i,+}\sigma_{C_j,-}+\underbrace{\sigma^2_{B,-}\otimes \sigma_{C_i,+}\sigma_{C_j,+}}_{0})\\
&=\sum_{i,j} \Omega_{i}\Omega_{j}(\sigma_{B,+}\sigma_{B,-}\otimes \sigma_{C_i,-}\sigma_{C_j,+}+\sigma_{B,-}\sigma_{B,+}\otimes \sigma_{C_i,+}\sigma_{C_j,-}),\\
\end{aligned}
\end{equation}
where we have used the fact that $\sigma^2_{B,\pm}\ket{0}_B=\sigma^2_{B,\pm}\ket{1}_B=0$, hence terms containing $\sigma^2_{B,\pm}$ contribute nothing and can thereby be dropped. Similarly,
\begin{equation}
	\begin{aligned}
H_1^3&=\sum_{i,j} \Omega_{i}\Omega_{j}(\sigma_{B,+}\sigma_{B,-}\otimes \sigma_{C_i,-}\sigma_{C_j,+}+\sigma_{B,-}\sigma_{B,+}\otimes \sigma_{C_i,+}\sigma_{C_j,-}) \sum_{k}\Omega_{k}(\sigma_{B,+}\otimes \sigma_{C_k,-}+\sigma_{B,-}\otimes \sigma_{C_k,+})\\
&=\sum_{i,j,k} \Omega_{i}\Omega_{j}\Omega_{k}(\underbrace{\sigma_{B,+}\sigma_{B,-}\sigma_{B,+}}_{\sigma_{B,+}}\otimes\sigma_{C_i,-} \sigma_{C_j,+}\sigma_{C_k,-}+\underbrace{\sigma_{B,-}\sigma_{B,+}\sigma_{B,-}}_{\sigma_{B,-}}\otimes \sigma_{C_i,+}\sigma_{C_j,-}\sigma_{C_k,+})\\
&=\sum_{i,j,k} \Omega_{i}\Omega_{j}\Omega_{k}(\sigma_{B,+}\otimes\sigma_{C_i,-} \sigma_{C_j,+}\sigma_{C_k,-}+\sigma_{B,-}\otimes \sigma_{C_i,+}\sigma_{C_j,-}\sigma_{C_k,+}),
\end{aligned}
\end{equation}
\begin{equation}
	\begin{aligned}
H_1^4&=\sum_{i,j,k,l} \Omega_{i}\Omega_{j}\Omega_{k}\Omega_{l}(\sigma_{B,+}\otimes\sigma_{C_i,-} \sigma_{C_j,+}\sigma_{C_k,-}+\sigma_{B,-}\otimes \sigma_{C_i,+}\sigma_{C_j,-}\sigma_{C_k,+})(\sigma_{B,+}\otimes \sigma_{C_l,-}+\sigma_{B,-}\otimes \sigma_{C_l,+})\\
&=\sum_{i,j,k,l} \Omega_{i}\Omega_{j}\Omega_{k}\Omega_{l} (\sigma_{B,+}\sigma_{B,-}\otimes\sigma_{C_i,-} \sigma_{C_j,+}\sigma_{C_k,-}\sigma_{C_l,+}+\sigma_{B,-}\sigma_{B,+}\otimes \sigma_{C_i,+}\sigma_{C_j,-}\sigma_{C_k,+} \sigma_{l,-}).
\end{aligned}
\end{equation}
From the above patterns of the power of $H_1$, one can obtain a general expression of $H_1^{2n}$ and $H_1^{2n+1}$. On the other hand, if we assume all the new qubits C$_i$ are initially at the ground state, then when $H_1^{2n}$ or $H_1^{2n+1}$ is applied to $\rho_{AB}\otimes\ket{00\dots00}_{C_1C_2 \dots C_n}\bra{00\dots00}_{C_1C_2 \dots C_n}$, all terms of $H_1^{2n}$ or $H_1^{2n+1}$ that contain $\sigma_{C_i,-}$ to its rightest side vanish since $\sigma_{C_i,-}\ket{0}=0$. This means that by making use of 
\begin{equation}
\sum_{i,j}\Omega_i \Omega_j\sigma_{C_i,-}\sigma_{C_j,+}\ket{00\dots00}_{C_1 C_2 \dots C_n}=\sum_{i=1}^n{\Omega^2_i}\sigma_{C_i,-}\sigma_{C_i,+}\ket{00\dots00}_{C_1 C_2 \dots C_n}=\sum_{i=1}^n{\Omega^2_i}\ket{00\dots00}_{C_1 C_2 \dots C_n},
\end{equation}
we can obtain an \textit{effective} ``Hamiltonian'' $H_L$ when $H_1$ is applied to $\ket{00\dots00}_{C_1C_2 \dots C_n}$:
\begin{equation}
	\begin{aligned}
		H_L&\cong\sum_i \Omega_i(\sigma_{B,-}\otimes \sigma_{C_i,+})\\
		H_L^2&\cong\sum_i \Omega_i^2 (\sigma_{B,+}\sigma_{B,-}\otimes \sigma_{C_i,-}\sigma_{C_i,+}) \cong\sum_i \Omega_i^2 (\sigma_{B,+}\sigma_{B,-}\otimes \mathbbm{1}_{C_i}) \\
		H_L^3&\cong\sum_{i,j}\Omega_i\Omega^2_j(\sigma_{B,-}\otimes \sigma_{C_i,+}\sigma_{C_j,-}\sigma_{C_j,+})\cong\sum_{i,j}\Omega_i\Omega^2_j(\sigma_{B,-}\otimes \sigma_{C_i,+})\\
		H_L^4&\cong\sum_{i,j}\Omega^2_i\Omega^2_j(\sigma_{B,+}\sigma_{B,-}\otimes \sigma_{C_i,-}\sigma_{C_i,+}\sigma_{C_j,-}\sigma_{C_j,+})\cong\sum_{i,j}\Omega^2_i\Omega^2_j(\sigma_{B,+}\sigma_{B,-}\otimes \mathbbm{1}_{C_i})\\
		&\vdots
	\end{aligned}
\end{equation}
where $\cong$ means ``the left- and right-hand side are effectively equal in the specific setup we are interested in''.

Note that $H_L$ and its powers are no longer Hermitian and should be regarded as an effective operator. Similarly, we can obtain an effective ``Hamiltonian'' $H_R$ when $H_1$ is applied to  $\bra{00\dots00}_{C_1C_2 \dots C_n}$:
\begin{equation}
	\begin{aligned}
		H_R&\cong\sum_i \Omega_i(\sigma_{B,+}\otimes \sigma_{C_i,-})\\
		H_R^2&\cong\sum_i \Omega_i^2 (\sigma_{B,+}\sigma_{B,-}\otimes \sigma_{C_i,-}\sigma_{C_i,+})\cong\sum_i \Omega_i^2 (\sigma_{B,+}\sigma_{B,-}\otimes \mathbbm{1}_{C_i})\\
		H_R^3&\cong\sum_{i,j}\Omega_i\Omega^2_j(\sigma_{B,+}\otimes \sigma_{C_i,-}\sigma_{C_j,+}\sigma_{C_j,-})\cong\sum_{i,j}\Omega_i\Omega^2_j(\sigma_{B,+}\otimes \sigma_{C_i,-})\\
		H_R^4&\cong\sum_{i,j}\Omega^2_i\Omega^2_j(\sigma_{B,+}\sigma_{B,-}\otimes \sigma_{C_i,-}\sigma_{C_i,+}\sigma_{C_j,-}\sigma_{C_j,+})\cong\sum_{i,j}\Omega^2_i\Omega^2_j(\sigma_{B,+}\sigma_{B,-}\otimes \mathbbm{1}_{C_i})\\
		&\vdots
	\end{aligned}
\end{equation}

Define $\mathbbm{H}_L=\mathbbm{1}_A\otimes H_L$ and $\mathbbm{H}_R=\mathbbm{1}_A\otimes H_R$, and denote $\rho=\rho_{ABC_1 C_2 \dots C_n}$, then
\begin{equation}\label{eq:eqd8}
\begin{aligned}
	\rho^\prime_{ABC_k}&=\operatorname{Tr}_{\text{all other C's}}\left[\exp(-i H_1 t) \rho\exp(i H_1 t)\right]\\
	&=\operatorname{Tr}_{\text{all other C's}}\left[\left(\mathbbm{1}-i H_1 t-\frac{1}{2!} H_1^{2} t^{2}+\frac{1}{3!} i H_1^{3} t^{3}+\ldots\right)\rho\left(\mathbbm{1}+i H_1 t-\frac{1}{2!} H_1^{2} t^{2}-\frac{1}{3!} i H_1^{3} t^{3}+\ldots\right)\right]\\
	&=\operatorname{Tr}_{\text{all other C's}}\left[\left(\mathbbm{1}-i \mathbbm{H}_L t-\frac{1}{2!} \mathbbm{H}_L^{2} t^{2}+\frac{1}{3!} i \mathbbm{H}_L^{3} t^{3}+\ldots\right)\rho\left(\mathbbm{1}+i \mathbbm{H}_R t-\frac{1}{2!} \mathbbm{H}_R^{2} t^{2}-\frac{1}{3!} i \mathbbm{H}_R^{3} t^{3}+\ldots\right)\right]\\
	&=\operatorname{Tr}_{\text{all other C's}}\left[\sum_{i,j=1}^{n}\left(\mathbbm{1}+\tilde{\sigma}_{-+}+ \tilde{\sigma}_{i,+}+\tilde{\sigma}_{i,-}+\tilde{\sigma}_{+-}\right) \rho\left(\mathbbm{1}+\tilde{\sigma}_{-+}^{*}+ \tilde{\sigma}_{j,+}^{*}+\tilde{\sigma}_{j,-}^{*}+\tilde{\sigma}_{+-}^{*}\right)\right]\\
	&=\left(\mathbbm{1}+\tilde{\sigma}_{-+}+ \sum_{i=1}^n \tilde{\sigma}_{i,+}+\sum_{i=1}^n\tilde{\sigma}_{i,-}+\tilde{\sigma}_{+-}\right) \underbrace{\rho_{ABC_{k}}}_{\tilde{\rho}}\left(\mathbbm{1}+\tilde{\sigma}_{-+}^{*}+ \sum_{i=1}^n\tilde{\sigma}_{i,+}^{*}+\sum_{i=1}^n\tilde{\sigma}_{i,-}^{*}+\tilde{\sigma}_{+-}^{*}\right)\\
	&=\tilde{\rho}+\tilde{\sigma}_{-+}\tilde{\rho}+\tilde{\rho} \tilde{\sigma}^*_{-+}+\tilde{\sigma}_{-+} \tilde{\rho}\tilde{\sigma}^*_{-+}+\tilde{\sigma}_{k,+}\tilde{\rho}+\tilde{\rho}\tilde{\sigma}^*_{k,-}+\tilde{\sigma}_{-+}\tilde{\rho}\tilde{\sigma}^*_{k,-}+\tilde{\sigma}_{k,+}\tilde{\rho}\tilde{\sigma}^*_{-+}\\&+\operatorname{Tr}_\text{all other C's except $C_k$}\left[\sum_{i=1}^n\tilde{\sigma}_{i,+}\tilde{\rho}\tilde{\sigma}_{i,-}\right]+\text{vanishing terms},
	\end{aligned}
\end{equation}
where we have denoted
\begin{equation}
\begin{aligned}
	\tilde{\sigma}_{-+}&=\sum_{i=1}^\infty\left[\frac{(-i t)^{2 i}}{2 i!}\left(\sum_{j=1}^{n} \Omega_{j}^{2}\right)^{i}\right]\sum_{k=1}^n \left(\mathbbm{1}_A\otimes \sigma_{B,+}\sigma_{B,-}\otimes\sigma_{C_k,-} \sigma_{C_k,+}\right)\cong\sum_{k=1}^n(\cos(t |\vec{\Omega}|)-1)\mathbbm{1}_A\otimes \sigma_{B,+}\sigma_{B,-}\otimes\mathbbm{1}_{C_k}\\
	\tilde{\sigma}_{+-}&=\sum_{i=1}^\infty\left[\frac{(-i t)^{2 i}}{2 i!}\left(\sum_{j=1}^{n} \Omega_{j}^{2}\right)^{i}\right]\sum_{k=1}^n \left(\mathbbm{1}_A\otimes \sigma_{B,-}\sigma_{B,+}\otimes\sigma_{C_k,+} \sigma_{C_k,-}\right)\cong\sum_{k=1}^n(\cos(t |\vec{\Omega}|)-1)\mathbbm{1}_A\otimes \sigma_{B,-}\sigma_{B,+}\otimes\mathbbm{1}_{C_k}\\
	\tilde{\sigma}_{k,-}&=\sum_{i=1}^\infty \left[\frac{(-i t)^{2 i}}{(2 i-1)!} \Omega_{k}\left(\sum_{j=1}^{n} \Omega_{j}^{2}\right)^{i-1}\right]\mathbbm{1}_A\otimes \sigma_{B,+}\otimes\sigma_{C_k,-}=-i \frac{\Omega_{k}}{|\vec{\Omega}|} \sin (t|\vec{\Omega}|)\mathbbm{1}_A\otimes \sigma_{B,+}\otimes\sigma_{C_k,-}\\
	\tilde{\sigma}_{k,+}&=\sum_{i=1}^\infty \left[\frac{(-i t)^{2 i-1}}{(2 i-1)!} \Omega_{k}\left(\sum_{j=1}^{n} \Omega_{j}^{2}\right)^{i-1}\right]\mathbbm{1}_A\otimes \sigma_{B,-}\otimes\sigma_{C_k,+}=-i \frac{\Omega_{k}}{|\vec{\Omega}|} \sin (t|\vec{\Omega}|)\mathbbm{1}_A\otimes \sigma_{B,-}\otimes\sigma_{C_k,+}.
	\end{aligned}
\end{equation} 
Note that in Eq.\,(\ref{eq:eqd8}), $\tilde{\sigma}_{-+}\tilde{\rho}$ is a sloppy notation and should be understood as $\Tr_\text{all other C's except $C_k$} [\tilde{\sigma}_{-+}\tilde{\rho}]$, and similarly for other terms containing $\tilde{\sigma}_{-+}$ and $\tilde{\sigma}^*_{-+}$. The key in the derivation in the last line of Eq.\,(\ref{eq:eqd8}) is to realize the non-vanishing terms of
\begin{equation}
	(\mathbbm{1}+\sigma_{C_k,+} +\sigma_{C_k,-}+\sigma_{C_k,+}\sigma_{C_k,-}+\sigma_{C_k,-}\sigma_{C_k,+})\rho_{C_k}(\mathbbm{1}+\sigma_{C_k,+} +\sigma_{C_k,-}+\sigma_{C_k,+}\sigma_{C_k,-}++\sigma_{C_k,-}\sigma_{C_k,+}),
\end{equation}
where $\rho_{C_k}=\ket{0}\bra{0}$.

\begin{table}[!b]
  \centering
  \caption{Recurrence relations of the reduced density matrices $\rho_{AB}$, $\rho_{BC_k}$, $\rho_{AC_k}$ and $\rho_{C_jC_k}$ of many-to-one collision model where all new qubits $C_1, C_2,\dots, C_n$ are initially at $\ket{0}$ and $\rho_{AB}$ has the X structure. Note that here, $\rho_{ij}=[\rho_{AB}]_{ij}$ whereas $\rho^\prime_{ij}=[\rho^\prime_{XY}]_{ij}$ where $XY=AB, AC_k, BC_k$ or $C_jC_k$.}
    \begin{tabular}{ccccc}
    \toprule
    $[\rho^\prime_{XY}]_{ij}$&$\rho^\prime_{AB}$&$\rho^\prime_{BC_k}$ & $\rho^\prime_{AC_k}$    & $\rho^\prime_{C_jC_k}$ \\
    \midrule
        &&&&\\[-1em]
    $\rho^\prime_{11}$&$\rho_{11}\cos^2(t|\vec{\Omega}|)$     &  0     &   $\rho_{11}\frac{\Omega^2_k}{|\vec{\Omega}|^2}\sin^2(t|\vec{\Omega}_k|)$ &   0\\ 
    &&&&\\[-1em]
  $ \rho^\prime_{22}$&$\rho_{22}+\rho_{11}\sin^2(t|\vec{\Omega}|)$     &   $(\rho_{11}+\rho_{33})\cos^2(t |\vec{\Omega})$    &   $\rho_{11}+\rho_{22}-\rho_{11}\frac{\Omega_{k}^{2}}{|\vec{\Omega}|^{2}} \sin ^{2}(t|\vec{\Omega}|)$   & $\left(\rho_{11}+\rho_{33}\right) \frac{\Omega_{j}^{2}}{|\vec{\Omega}|^{2}} \sin ^{2}(t|\vec{\Omega}|)$ \\
   &&&&\\[-1em]
   $ \rho^\prime_{33}$&$\rho_{33}\cos^2(t|\vec{\Omega}|)$     & $(\rho_{11}+\rho_{33})\frac{\Omega^2_{C_k}}{|\vec{\Omega}|^2}\sin^2(t |\vec{\Omega})$     &    $\rho_{33}\frac{\Omega^2_k}{|\vec{\Omega}|^2}\sin^2(t|\vec{\Omega}_k|)$   &$\left(\rho_{11}+\rho_{33}\right) \frac{\Omega_{k}^{2}}{|\vec{\Omega}|^{2}} \sin ^{2}(t|\vec{\Omega}|)$  \\
    &&&&\\[-1em]
    $\rho^\prime_{44}$&$\rho_{44}+\rho_{33}\sin^2(t|\vec{\Omega}|)$     &  $1-\rho^\prime_{22}-\rho^\prime_{33}$ &   $\rho_{33}+\rho_{44}-\rho_{33} \frac{\Omega_{k}^{2}}{|\vec{\Omega}|^{2}} \sin ^{2}(t|\vec{\Omega}|)$    & $1-\rho^\prime_{22}-\rho^\prime_{33}$\\
     &&&&\\[-1em]
    $\rho^\prime_{23}$&$\rho_{23}\cos(t|\vec{\Omega}|)$     &    $\frac{i \Omega_k (\rho_{11}+\rho_{33})\sin(2t|\vec{\Omega}|)}{2|\vec{\Omega}|}$   &    $i \rho_{23} \frac{\Omega_{k}}{|\vec{\Omega}|} \sin (t|\vec{\Omega}|)$   & $\left(\rho_{11}+\rho_{33}\right) \frac{\Omega_{j} \Omega_{k}}{|\vec{\Omega}|^{2}} \sin ^{2}(t|\vec{\Omega}|)$ \\
     &&&&\\[-1em]
   $ \rho^\prime_{14}$&$\rho_{14}\cos(t|\vec{\Omega}|)$     &    0   &    $-i \rho_{14} \frac{\Omega_{k}}{|\vec{\Omega}|} \sin (t|\vec{\Omega}|)$   & 0 \\
   \bottomrule
    \end{tabular}%
  \label{tab:appendixtable1}%
\end{table}

To proceed, suppose that $\rho_{AB}$ is initially at the X state, such that
\begin{equation}
	\rho_{ABC_k}=\begin{pmatrix}
		\rho_{11}&0&0&\rho_{14}\\
		0&\rho_{22}&\rho_{23}&0\\
		0&\rho_{32}&\rho_{33}&0\\
		\rho_{41}&0&0&\rho_{44}
	\end{pmatrix}_{AB}\otimes \begin{pmatrix}
		0&0\\
		0&1
	\end{pmatrix}_{C_k}.
\end{equation}
Plugging in $\rho_{ABC_k}$ into Eq.\,(\ref{eq:eqd8}), one obtains $\rho^\prime_{ABC_k}$, from which one can calculate $\rho^\prime_{AB}$, $\rho^\prime_{BC_k}$ and $\rho^\prime_{AC_k}$ by taking partial traces. Regarding $\rho^\prime_{C_jC_k}$, one can calculate it similarly as in Eq.\,(\ref{eq:eqd8}) by carefully collecting non-vanishing terms,
\begin{equation}
\begin{aligned}
	\rho^\prime_{ABC_j C_k}&=\left(\mathbbm{1}+\tilde{\sigma}_{+-}+\tilde{\sigma}_{-+}+\sum_{i=1}^{n} \tilde{\sigma}_{i,+}+\sum_{i=1}^{n} \tilde{\sigma}_{i,-}\right) \underbrace{\rho_{A B C_j C_{k}}}_{\tilde{\rho}}\left(\mathbbm{1}+\tilde{\sigma}_{+-}^*+\tilde{\sigma}_{-+}^*+\sum_{i=1}^{n} \tilde{\sigma}_{i,+}^{*}+\sum_{i=1}^{n} \tilde{\sigma}_{i,-}^{*}\right)\\
	&=\tilde{\rho}+\tilde{\sigma}_{-+} \tilde{\rho}+\tilde{\rho} \tilde{\sigma}_{-+}^{*}+\tilde{\sigma}_{-+} \tilde{\rho} \tilde{\sigma}_{-+}^{*}+\sum_{i=j,k}\left(\tilde{\sigma}_{i,+} \tilde{\rho}+\tilde{\rho} \tilde{\sigma}_{i,-}^{*}+\tilde{\sigma}_{-+} \tilde{\rho} \tilde{\sigma}_{i,-}^{*}+\tilde{\sigma}_{i,+} \tilde{\rho} \tilde{\sigma}_{-+}^{*}\right)\\
	&+\tilde{\sigma}_{j,+}\tilde{\rho}\sigma^*_{k,-}+\tilde{\sigma}_{k,+}\tilde{\rho}\tilde{\sigma}^*_{j,-}+\Tr_\text{all C's except $C_j$ and $C_k$} \left[\sum_{i=1}^n \tilde{\sigma}_{i,+}\tilde{\rho}\tilde{\sigma}^*_{i,-}\right]+\text{vanishing terms},
	\end{aligned}
\end{equation}
where 
\begin{equation}
	\rho_{ABC_jC_k}=\begin{pmatrix}
		\rho_{11}&0&0&\rho_{14}\\
		0&\rho_{22}&\rho_{23}&0\\
		0&\rho_{32}&\rho_{33}&0\\
		\rho_{41}&0&0&\rho_{44}
	\end{pmatrix}_{AB}\otimes \begin{pmatrix}
		0&0\\
		0&1
	\end{pmatrix}_{C_j}\otimes \begin{pmatrix}
		0&0\\
		0&1
	\end{pmatrix}_{C_k}.
\end{equation}
As before, $\tilde{\sigma}_{-+}\tilde{\rho}$ is a sloppy notation and should be understood as $\Tr_\text{all other C's except $C_k$ and $C_j$} [\tilde{\sigma}_{-+}\tilde{\rho}]$, and similarly for other terms containing $\tilde{\sigma}_{-+}$ and $\tilde{\sigma}^*_{-+}$. Note that because A is initially correlated with B, one cannot just calculate $\rho^\prime_{AC_jC_k}$ or $\rho^\prime_{BC_jC_k}$. Instead, one has to calculate $\rho^\prime_{ABC_jC_k}$ before tracing out A and B to obtain $\rho^\prime_{C_j C_k}$.
 In Table \ref{tab:appendixtable1} we summarize the results (in the rotating frame where the contribution from $H_2$ is voided).

 \end{widetext}

\bibliography{apssamp4}

\end{document}